\documentclass[letterpaper,twocolumn,10pt]{article}
\usepackage{usenix}
\usepackage{cite}
\usepackage{amsmath,amssymb,amsfonts}
\usepackage{graphicx}
\usepackage{textcomp}
\usepackage{xcolor}
\def\BibTeX{{\rm B\kern-.05em{\sc i\kern-.025em b}\kern-.08em
    T\kern-.1667em\lower.7ex\hbox{E}\kern-.125emX}}
\usepackage{hyperref}  
\usepackage{array}
\usepackage[caption=false,font=footnotesize,labelfont=rm,textfont=rm]{subfig}
\usepackage{textcomp}
\usepackage{stfloats}
\usepackage{url}
\usepackage{verbatim}
\usepackage{graphicx}
\usepackage{cite}
\usepackage{gensymb}
\usepackage{multirow}
\usepackage{booktabs}
\usepackage{colortbl}
\usepackage{makecell}
\usepackage{enumitem}
\usepackage{xcolor}
\usepackage{tikz}
\usepackage{amsmath}
\usepackage[linesnumbered,ruled,vlined,algo2e]{algorithm2e}
\usepackage[ruled, vlined, linesnumbered]{algorithm2e}
\usepackage{algorithm}
\usepackage{algpseudocode}
\usepackage{multirow}
\usepackage{color}
\usepackage{filecontents}
\usepackage{subfig,graphicx}
\usepackage{xspace}

\newcommand\SystemName{\textsc{EchoMask}\xspace}

\newcommand\cparagraph[1]{\vspace{1.2mm}\noindent \textbf{#1.}}

\begin{document}
\author{Zhiyuan Ning\thanks{Northwest University} \and Zhanyong Tang\thanks{Northwest University} \and Xiaojiang Chen\thanks{Northwest University} \and Zheng Wang\thanks{University of Leeds}}

\title{Before the Mic: Physical-Layer Voiceprint Anonymization with Acoustic Metamaterials}
\maketitle

\begin{abstract}
%
Voiceprints are widely used for authentication; however, they are easily captured in public settings and cannot be revoked once leaked. Existing anonymization systems operate inside recording devices, which makes them ineffective when microphones or software are untrusted, as in conference rooms, lecture halls, and interviews.  
We present \SystemName, the first practical physical-layer system for real-time voiceprint anonymization using acoustic metamaterials. By modifying sound waves before they reach the microphone, \SystemName prevents attackers from capturing clean voiceprints through compromised devices. Our design combines three key innovations: frequency-selective interference to disrupt voiceprint features while preserving speech intelligibility, an acoustic-field model to ensure stability under speaker movement, and reconfigurable structures that create time-varying interference to prevent learning or canceling a fixed acoustic pattern.  
\SystemName is low-cost, power-free, and 3D-printable, requiring no machine learning, software support, or microphone modification. Experiments conducted across eight microphones in diverse environments demonstrate that \SystemName increases the Miss-match Rate, i.e., the fraction of failed voiceprint matching attempts, to over 90\%, while maintaining high speech intelligibility.
\end{abstract}

\section{Introduction}
\vspace{-2mm}
As a biometric trait, voiceprints are widely used for identity verification and secure access~\cite{Voiceprint1,Voiceprint2,Voiceprint3,Voiceprint4,Voiceprint5}. At the same time, they are easy to capture in many real-world settings: when speech is recorded by compromised or untrusted devices (e.g., microphones or recording infrastructure), an adversary can extract and reuse voiceprints for impersonation, leading to privacy breaches and financial fraud.

This risk is especially high in public settings such as talks, meetings, or interviews, where users must rely on third-party microphones and open networks~\cite{attack1,attack2,attack3,attack4,attack5}. This differs from voice data released in recorded videos or audio clips, where voices can be anonymized before publication. In such public settings, users have little control over how their speech is captured and stored in real time. Protecting voiceprints at the moment of capture is therefore critical. Real-time voiceprint anonymization creates an immediate privacy barrier by altering identity-bearing features before the audio is stored or transmitted~\cite{NTU-NPU,saic}.

Existing solutions are either software- or hardware-based~\cite{vsmask,micpro,enkidu,vcloak}. Software methods add perturbations or transformations to the speech signal after it is captured by the microphone~\cite{vsmask,enkidu,vcloak}. This design assumes that the recording device is trustworthy. In practice, however, if an attacker controls the device or gains access to the microphone (e.g., via a compromised smartphone), they can record the raw speech signal before anonymization is applied, which defeats the protection. Hardware-based approaches move anonymization into the microphone itself~\cite{micpro}, which offers stronger security. However, they often rely on specific speech codecs, limiting their generalizability across different devices. Moreover, hardware-based solutions require microphone encoding, resulting in device-specific characteristics that make them difficult to adapt to public scenarios that require third-party microphones (such as public speeches). These limits point to a deeper problem: both software and hardware solutions operate inside the device. Once speech has entered the microphone, it is already too late to protect privacy.

In this work, we show for the first time that acoustic metamaterials can provide a physical-layer solution for real-time voiceprint anonymization. Acoustic metamaterials are engineered structures that control sound propagation by shaping wavefronts and modulating amplitude and phase in selected frequency bands~\cite{r11,r12,r14,r15}. They have previously been used for noise control, acoustic filtering, blocking ultrasound attacks~\cite{mitigating}, and guiding weak sound signals~\cite{nature}.  Our work repurposes these capabilities to protect speaker identity. By placing a passive metamaterial outside the microphone, we distort identity-bearing components of speech before capture while preserving intelligibility. The compact, power-free design can be attached to off-the-shelf microphones, making it suitable for public and shared recording environments.

However, utilizing metamaterials for voiceprint anonymization in real-life settings requires overcoming three obstacles. First, voiceprint features and speech recognition rely on overlapping frequency bands, so naive filtering will destroy speech quality.  Second, in practical applications, dynamic changes in the speaker's posture, such as head movement, can alter the incident angle of sound waves, causing significant fluctuations in the acoustic pressure received by the metamaterial, making anonymization unstable.  Third, current metamaterials have fixed acoustic response patterns, but a static perturbation can be analyzed and potentially removed by an attacker, reducing long-term reliability.

\begin{figure}[t!]
    \centering
    \includegraphics[width=\linewidth]{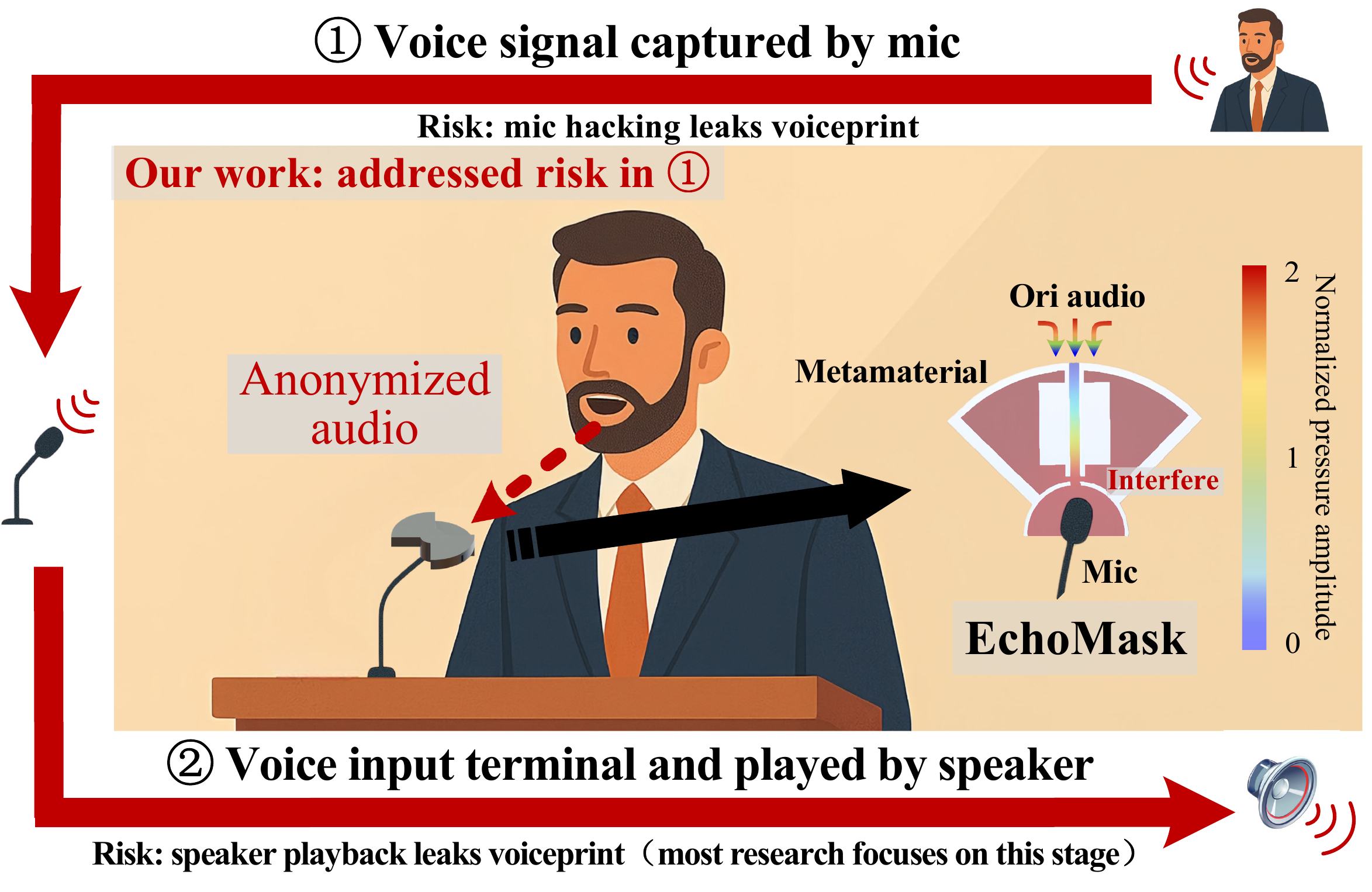}
    \vspace{-2mm}
    \caption{A deployment scenario of \SystemName for voiceprint anonymization.}
    \label{fig:examples}
\end{figure}

We present \SystemName, the first practical acoustic metamaterial system for real-time voiceprint anonymization. Fig.~\ref{fig:examples} depicts a deployment scenario of \SystemName.  Unlike prior anonymization solutions, \SystemName jointly co-designs frequency-selective interference, a dynamic acoustic-field model, and reconfigurable structures to achieve strong voiceprint disruption while preserving speech intelligibility and stability under user movement.

To this end, we first exploit the tolerance of speech recognition to distortions across frequency bands, allowing \SystemName to identify where strong voiceprint disruption can be applied with minimal loss of intelligibility.  We then build an acoustic-field model that captures how sound pressure changes with speaker movement. This model lets us compute the minimum number of metamaterial units and their optimal orientations to maintain coverage over changing sound directions.  Finally, we introduce reconfigurable, power-free structures that change in response to small user motions. This causes the interference pattern to vary over time, preventing attackers from learning or canceling a fixed acoustic signature. 

We show that \SystemName can be produced using low-cost resin 3D printing and requires no external power. It relies on physical acoustics rather than machine learning models, avoiding training cost and runtime overhead. Because it works outside the microphone, it does not require any software support or changes to device hardware, making it easy to deploy on existing microphones.

We evaluate \SystemName using eight microphones from different vendors in varied environments. Experimental results show that \SystemName increases the  Miss-Match Rate to over 90\% across all devices without compromising the speech quality, showing that strong anonymization can be achieved without sacrificing usability. 

The main contributions of this paper are: 

\begin{itemize}[leftmargin=*]
\item The first real-time voiceprint anonymization system based on acoustic metamaterials for real-world deployment;
\item Addressing three key challenges that prevent metamaterials from working in dynamic speaking scenarios;
\item A low-cost and deployable defense built from 3D-printed acoustic metamaterials.
\end{itemize}

\section{Background} \label{chap:2}
\vspace{-2mm}
\subsection{Voiceprint Leakage}  
Voice interaction systems are widely used in authentication, virtual assistants, and online communication, making voice data an increasingly important biometric identifier. Voiceprints encode speaker-specific characteristics for recognition and verification, but their biometric nature also makes them a high-value target for attackers~\cite{easy,songbsab,navigating1,navigating2}.

Voiceprints are persistent and difficult to revoke, yet are routinely exposed during normal use. Attackers can capture speech from calls, meetings, interviews, or public talks and extract voiceprint features, enabling identity impersonation, authentication bypass, and related fraud~\cite{enkidu,vsmask}.

This risk is particularly high in public speaking settings, where users rely on third-party microphones and recording infrastructure, and sometimes untrusted networks~\cite{attack1,attack2,attack3,attack4,attack5}, and have little control over how their voice data is captured, processed, or stored in real time. This differs fundamentally from recorded audio or video releases, where voice data can be reviewed or anonymized prior to publication. Once leaked, voiceprints can be reused indefinitely across services, and recent advances in speech synthesis and voice cloning~\cite{SP2025} further amplify this threat. Protecting voiceprints at the point of capture is therefore essential, as post-hoc mitigation is largely ineffective.

\subsection{Threat Model}
\vspace{-2mm}
We consider an adversary whose goal is to obtain a target’s voiceprint for impersonation or identity abuse. The adversary may fully compromise the recording device and software stack (e.g., via a malicious application) and can access, store, and process all captured audio using state-of-the-art speaker recognition, voice cloning, and signal processing techniques~\cite{micpro,ivectorPLDA,Xvectors}. The adversary may collect arbitrary recordings produced by the protected microphone, including long-duration audio across multiple sessions.
This threat model reflects realistic settings such as public microphones, conference rooms, lecture halls, and outdoor speeches, where users must rely on third-party equipment that may be compromised in advance~\cite{micattack1,micattack2,micattack3,micattack4}. Accordingly, we do not assume that the microphone hardware, firmware, operating system, or recording software is trusted.

We assume the user speaks through a microphone equipped with \SystemName, which is physically attached to the microphone. The adversary cannot remove or tamper with the metamaterial during the speech, but is assumed to have full knowledge of the \SystemName's design and operating principles. Beyond this physical constraint, the adversary may observe and manipulate all digital audio output and attempt modeling- or compensation-based attacks that exploit \emph{static} voice perturbations. Our goal is to ensure that, even under full device compromise, the captured audio does not contain a reliable or reusable voiceprint. 
Finally, we treat speech intelligibility and usability as essential constraints. The protected audio must remain understandable to human listeners and usable by modern speech recognition systems (but not for user identification).

\subsection{Acoustic Metamaterials}
\SystemName leverages acoustic metamaterials as its physical-layer building block. Acoustic metamaterials are engineered structures designed to control how sound propagates in space, enabling precise manipulation of the phase and amplitude of sound waves through carefully designed subwavelength geometries~\cite{r11,r12,r14,r15}. By shaping acoustic responses in selected frequency bands, metamaterials enable fine-grained control over sound waves using passive resonant structures.

This capability provides a natural foundation for voiceprint anonymization. Unlike digital filters or software-based signal processing, acoustic metamaterials operate entirely at the physical layer: they modulate airborne sound directly before it is captured by the microphone, without analog-to-digital conversion, digital signal processing, or software support~\cite{mitigating,mie2,nature}. As a result, they eliminate risks associated with software compromise or microphone hijacking and enable anonymization even when the recording device is untrusted. Because metamaterials can be attached externally to the microphone, they also avoid the device-binding and compatibility issues of hardware modification approaches, while introducing no processing latency.

Despite the advantages, applying acoustic metamaterials to voiceprint anonymization faces multiple practical challenges. First, voiceprint features largely overlap with speech frequency bands, requiring interference designs that disrupt identity while preserving intelligibility. Second, speaker movement changes the angle of incoming sound waves, leading to variations in acoustic pressure that can destabilize the anonymization effect. Finally, conventional metamaterials exhibit fixed acoustic responses, which may be analyzed or adapted to by attackers, reducing long-term effectiveness. Addressing these challenges is essential for building a practical metamaterial-based voiceprint anonymization system.
\section{Our Approach} \label{chap:4}
\begin{figure}[!t]
\centering
\subfloat[]{
    \includegraphics[scale=0.137]{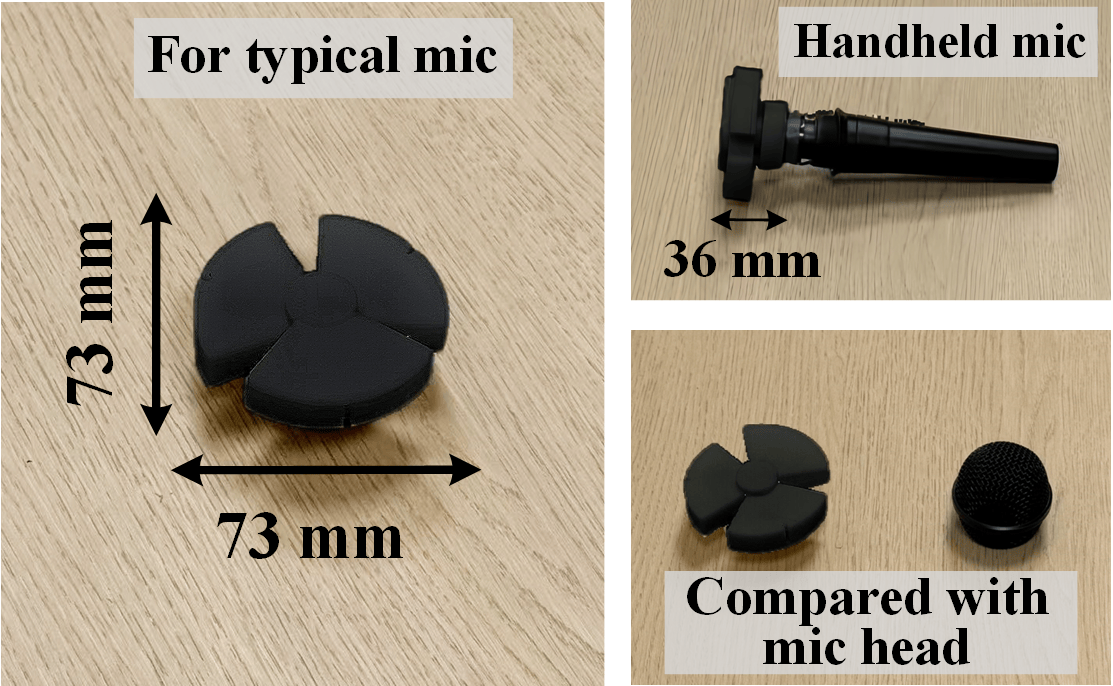}
    \label{typical}}
\hfill
\subfloat[]{
    \includegraphics[scale=0.137]{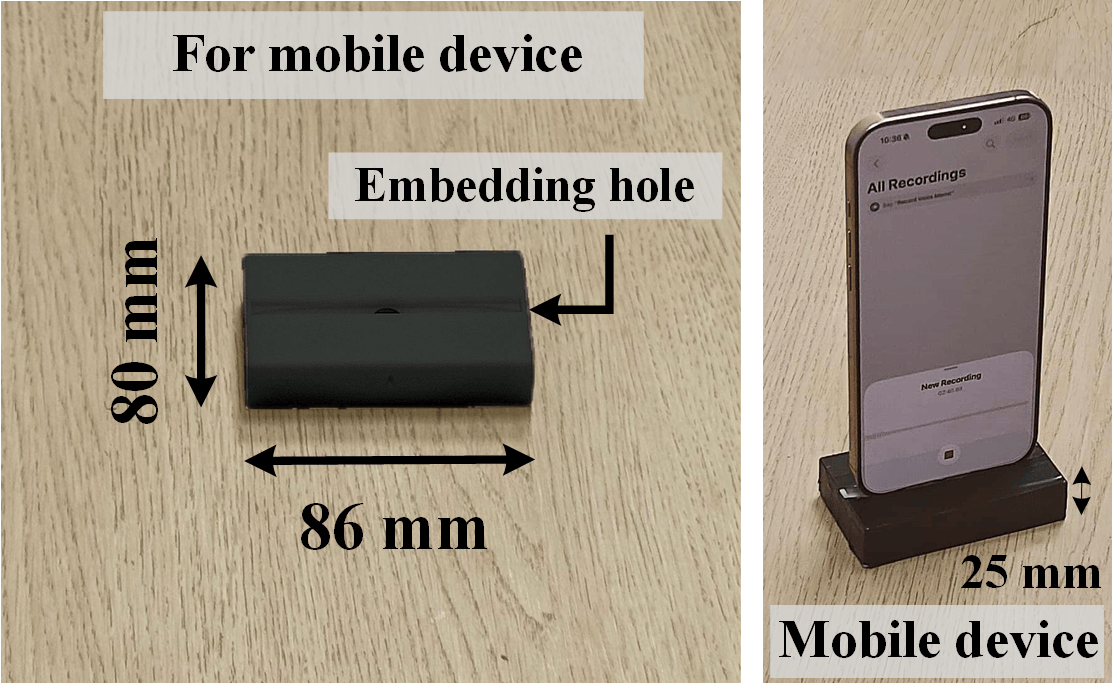}
    \label{smartphone}}
\caption{Prototypes for typical microphones (a) and mobile devices (b).}


\label{fig:prototype1}
\end{figure}

\vspace{-2mm}
\SystemName is a physical-layer, power-free voiceprint protection system that operates at the point of capture. By attaching directly to a microphone, it anonymizes the analog speech signal \emph{before} digitization, preventing voiceprint leakage regardless of downstream software, transmission, or storage practices. Fig.~\ref{fig:prototype1} shows a prototype of \SystemName and its attachment to a typical microphone and smartphone.

\subsection{Design Principles}
Our design is guided by three principles that address key limitations of prior voice anonymization approaches.

First, voice biometrics differ from traditional credentials in that identity cues are continuously exposed during normal use. Effective protection must therefore suppress speaker-specific information while preserving intelligibility and usability. Rather than applying arbitrary distortion, \SystemName separates the objectives of speaker recognition (i.e., user identification) and speech recognition (i.e., speech intelligibility), enabling targeted perturbation of identity-bearing features without degrading semantic content.

Second, capture-time protection must also remain effective under natural user behavior. As speakers move and change orientation, the acoustic path to the microphone varies, causing direction-sensitive defenses to fail silently. \SystemName treats angular robustness as a core requirement and designs physical interference that remains effective across realistic speaking angles.
Finally, long-term security requires resistance to modeling and compensation attacks~\cite{attacks1,attacks2}. Static passive perturbations can become predictable once observed. \SystemName addresses this by introducing controlled, power-free randomness into the acoustic response, producing time-varying interference patterns that are difficult to predict or invert, even with extensive recordings.

We implement the principles through three integrated components. It applies targeted low-frequency perturbation that exploits structural differences between speaker recognition and speech recognition (Sec.~\ref{sec:tlp}). It employs a dynamically stable multi-unit metamaterial layout that maintains strong interference across realistic speaking angles (Sec.~\ref{sec:layout}). Finally, it introduces passive randomization of the acoustic response to enhance long-term robustness against adaptive attacks (Sec.~\ref{sec:ril}). Together, these components enable robust, usable, and capture-time voiceprint anonymization without power, sensing, or software support.

\subsection{Targeted Low-Frequency Perturbation\label{sec:tlp}}

A key goal of \SystemName is to conceal voiceprint information while maintaining clear and intelligible speech. This is challenging because speaker identity and speech content are encoded in overlapping regions of the acoustic spectrum. If strong or wide-ranging perturbations are applied to disrupt speaker identity, they can also distort phonemes and significantly degrade speech recognition.

\subsubsection{Voiceprint perturbation band derivation}
To balance voiceprint concealment with speech intelligibility, we adopt a selective perturbation strategy. Effective protection must disrupt speaker identity extraction while preserving speech that remains intelligible to human listeners. To reason about this trade-off in a principled and measurable way, we analyze speaker recognition and speech recognition as proxies for these competing objectives: the former captures acoustic cues used for identity inference, while the latter reflects properties closely aligned with human speech understanding.

Our key insight is that these tasks rely on different acoustic structures. Speaker recognition depends strongly on low-frequency characteristics shaped by stable physiological factors, such as vocal tract length and vocal fold tension, which are highly discriminative across speakers~\cite{speakerrecognition1,speakerrecognition2}. In contrast, human speech intelligibility is primarily conveyed through phoneme articulation and temporal patterns in mid- and high-frequency bands, with substantial contextual redundancy. Modern automated speech recognition (ASR) systems are trained to recover linguistic content under noise and distortion and thus serve as a conservative proxy for human intelligibility rather than the protection target itself~\cite{asr1,asr2}. Consequently, selectively perturbing a narrow low-frequency band can significantly disrupt speaker identity while largely preserving speech understanding.

To identify this sensitive frequency region, we analyze the spectral characteristics of speaker recognition and speech recognition systems. Speaker recognition extracts a speaker embedding $\mathbf{e}$ from a time-frequency representation $X$ (e.g., Mel-frequency cepstral coefficients~\cite{MFCC}) and predicts the identity $\hat{s}$ that maximizes the posterior probability:
\begin{equation}
\hat{s} = \arg\max_{s} p(\mathbf{e} \mid s), \quad \mathbf{e} = f_{\text{SV}}(X),
\end{equation}
where $s$ denotes a candidate speaker. Prior studies show that low-frequency components, particularly those associated with the first formant (F1), carry a large fraction of identity-related information~\cite{formant1}. Even small perturbations in this region can cause substantial shifts in the embedding space, increasing intra-class variation or reducing inter-class separation, thereby leading to recognition errors. This sensitivity arises because F1 reflects vocal tract length and shape, which are key physiological cues of speaker identity.

In contrast, speech understanding, both by humans and by ASR systems, focuses on recovering \textit{what is being said}. This process can be expressed as predicting the most likely word sequence:
\begin{equation}
\hat{W} = \arg\max_W p(W \mid X).
\end{equation}
Phonetic and semantic information is largely encoded in higher-frequency structures, particularly the second and third formants (F2 and F3), as well as in temporal context~\cite{formant2}. If the speech representation is decomposed as
\begin{equation}
X = X_{\text{F1}} + X_{\text{F2,F3}},
\end{equation}
the decoding process can be approximated as
\begin{equation}
p(W \mid X) \approx p(W \mid X_{\text{F2,F3}}, \text{context}),
\end{equation}
indicating that linguistic content can be recovered even when low-frequency components are moderately degraded. Empirical studies and psychoacoustic evidence show that such low-frequency perturbations have a limited impact on human intelligibility~\cite{impact1}, which is consistent with the observed robustness of ASR systems to band-limited distortion.

Based on this analysis, we identify the low-frequency band adjacent to F1 as the most sensitive region for speaker recognition~\cite{formant1,formant2}. Combining spectral analysis with empirical observations, we select a perturbation band centered around 500 Hz, covering approximately 300-700 Hz. This band contains core speaker identity information while minimally affecting cues essential for human speech understanding, enabling targeted perturbations that weaken voiceprints while preserving speech clarity and intelligibility.

\subsubsection{Perturbation band design}\label{sec:band_design}
\begin{figure*}[!t]
  \begin{minipage}[!t]{0.48\linewidth}
\centering
\subfloat[]{
    \includegraphics[scale=0.07]{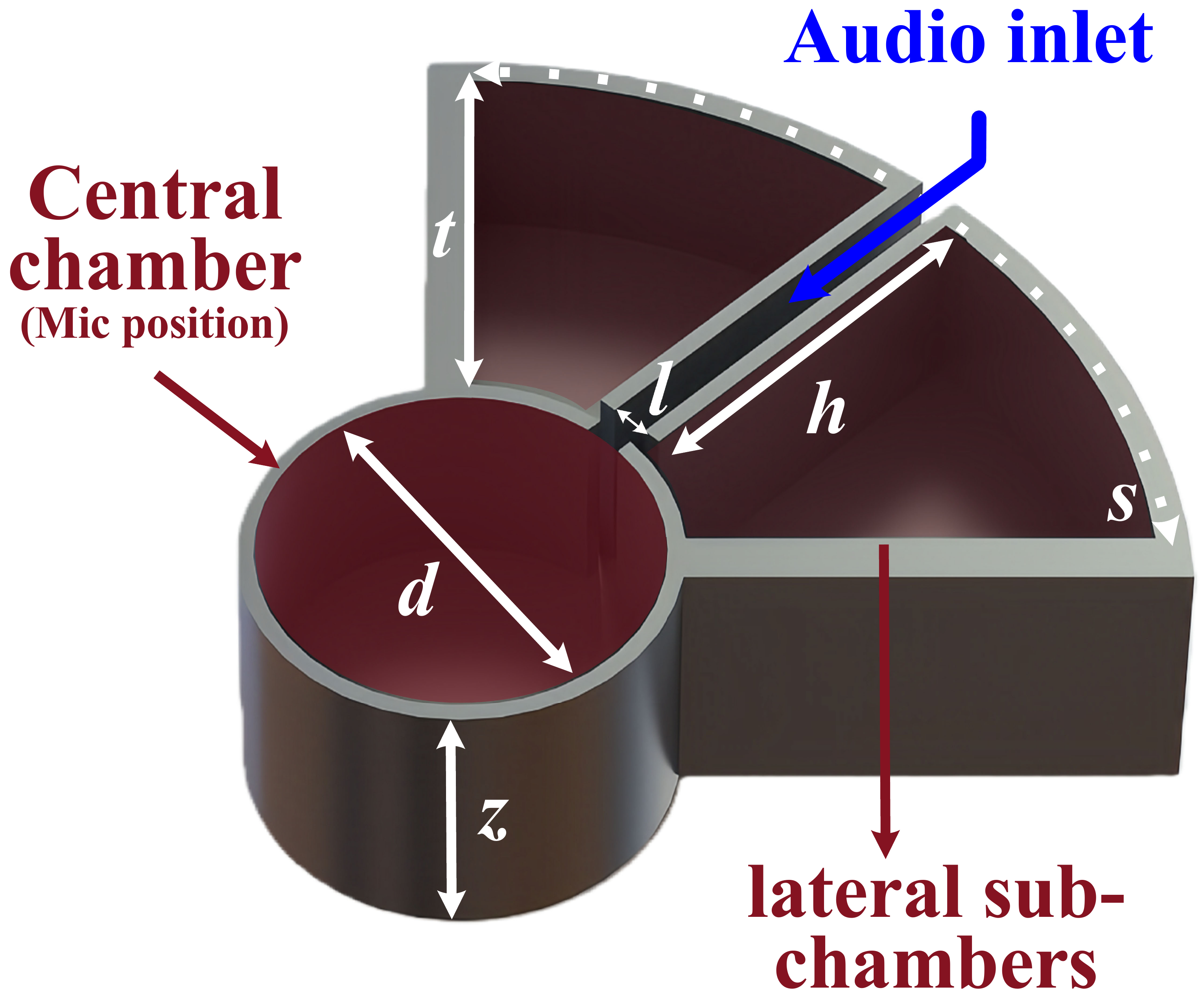}
    \label{structure1}}
\hfill
\subfloat[]{
    \includegraphics[scale=0.07]{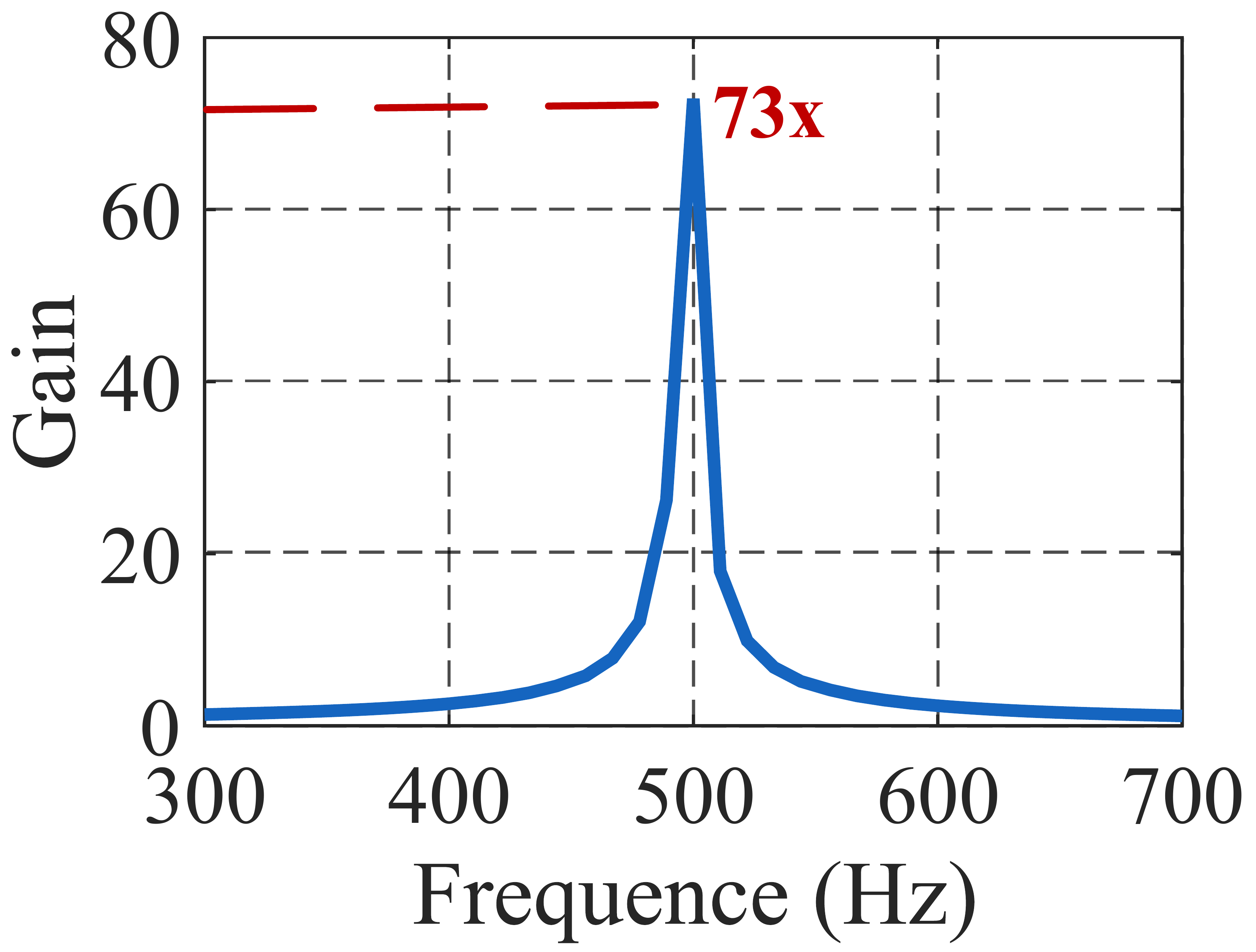}
    \label{structure2}}
\caption{(a) Voiceprint anonymization metamaterial and (b) its interference performance.}
\label{fig:metamaterial}
    \end{minipage}  
    \hfill
\begin{minipage}[!t]{0.48\linewidth}
\centering
\subfloat[]{ \includegraphics[scale=0.08]{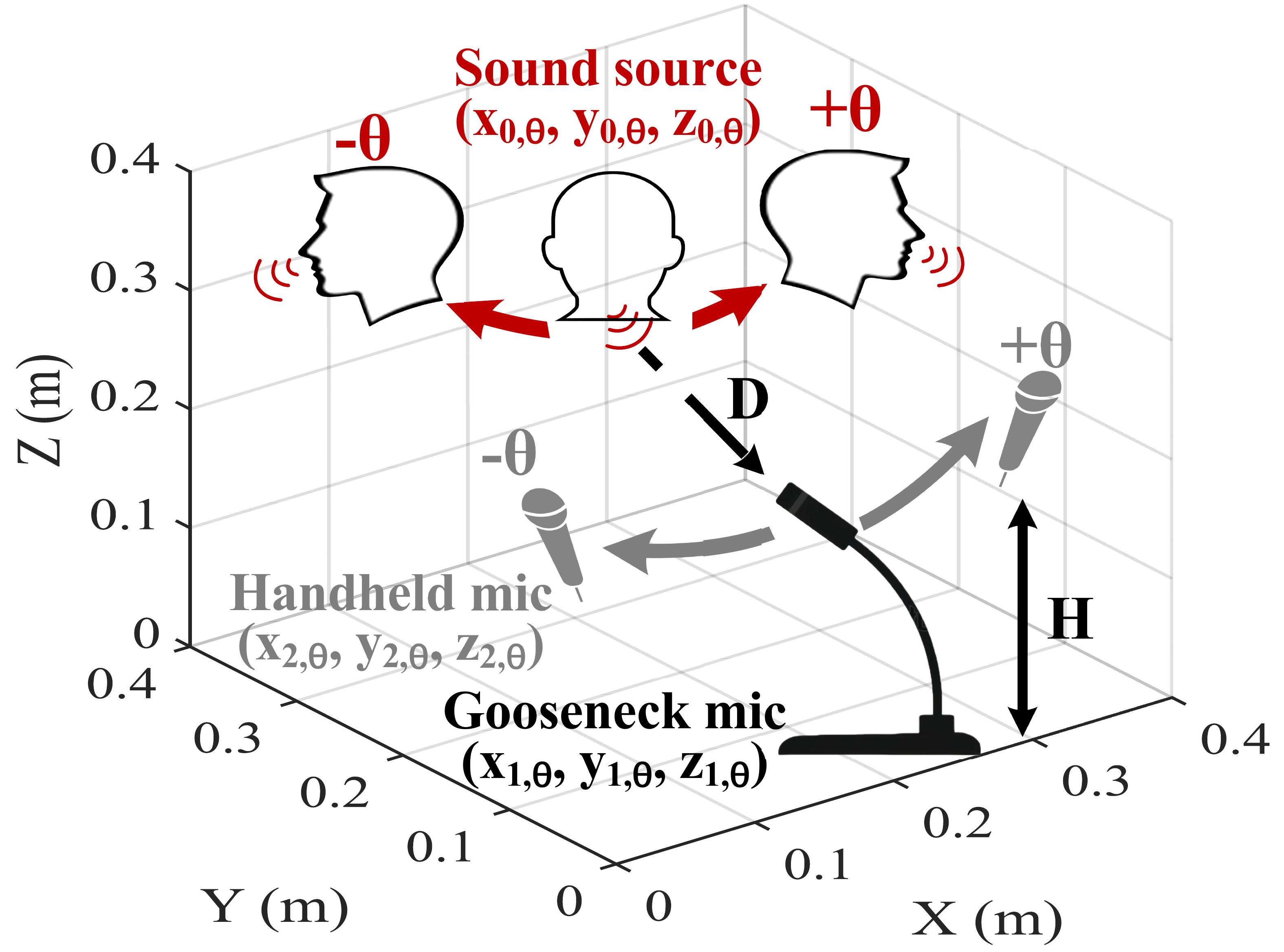}
  \label{model11}}
\hfill
\subfloat[]{
        \includegraphics[scale=0.065]{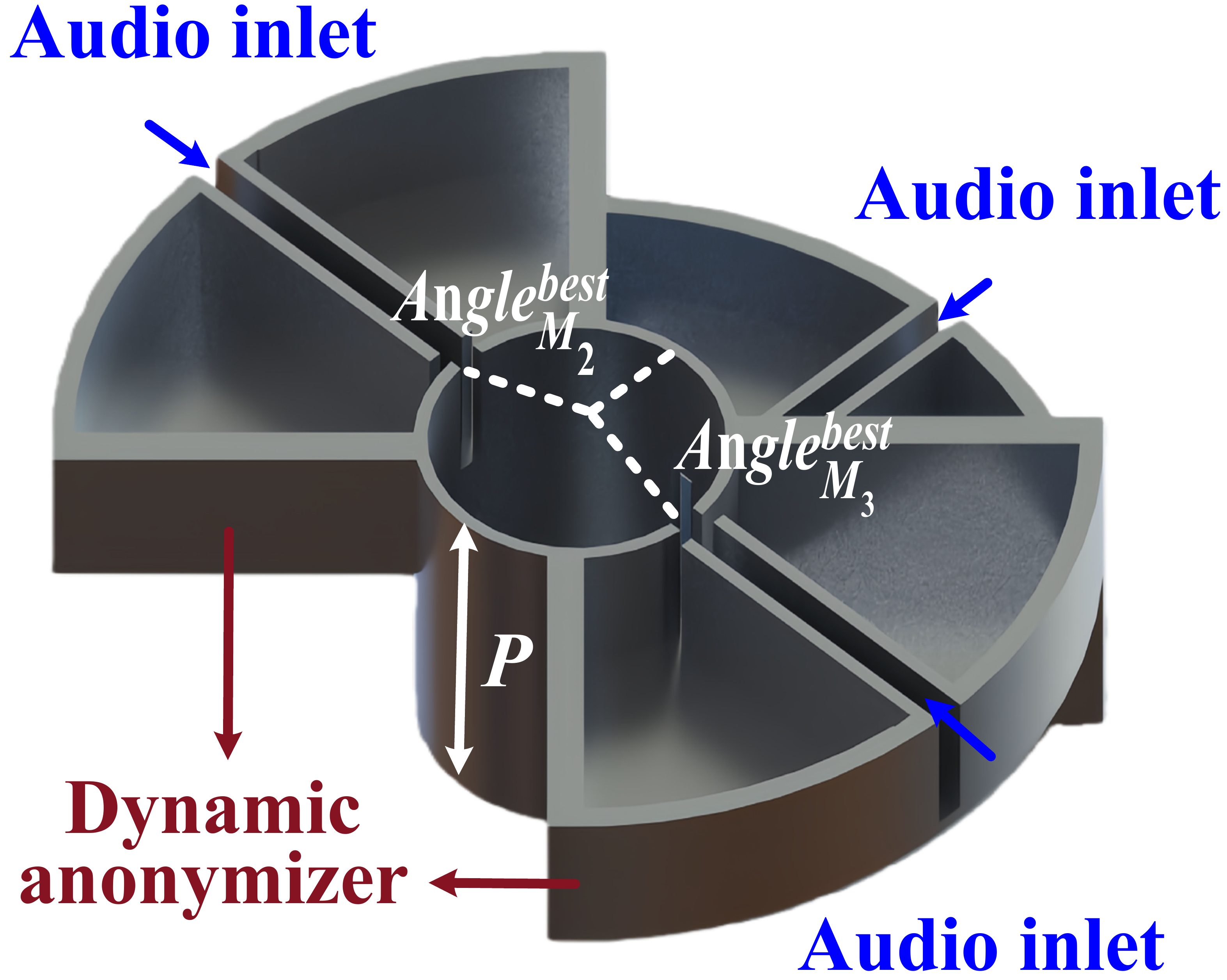}
  \label{model22}}
\caption{(a) Anonymization acoustic field model and (b) dynamic anonymization metamaterial.}
    \end{minipage}  
\end{figure*}


To precisely perturb voiceprints in the 300 - 700 Hz range, we design a compact acoustic metamaterial based on Mie resonators~\cite{mie,mie2}. At a high level, a Mie-resonator-based acoustic metamaterial uses small, subwavelength cavities to trap and amplify sound at specific frequencies. Although the structure itself is compact, these resonances allow it to strongly interact with low-frequency sound, in a way analogous to Mie resonances in electromagnetic scattering~\cite{nature}. This property makes such metamaterials well suited for passive, low-frequency manipulation of speech signals.

As shown in Fig.~\ref{structure1}, our design consists of a central cavity surrounded by multiple side cavities. The cavities support a strong monopole resonance that concentrates acoustic energy within a very small region of the structure. This localized energy amplification selectively disrupts speaker-specific acoustic cues, enabling effective anonymization in the target low-frequency band while leaving most speech content intact.

At resonance, the side cavities strongly enhance the structure's acoustic response. This effect can be described using a simple energy model:
\begin{equation}
W_{\mathrm{loc}}(\omega) = G(\omega)\, W_0(\omega),
\end{equation}
where $W_0(\omega)$ denotes the acoustic energy without the metamaterial, and $G(\omega)$ captures the resonance-induced amplification. Near the resonance frequency, this amplification can be approximated as
\begin{equation}
G(\omega) = \frac{A}{(\omega_0 - \omega)^2 + \gamma^2},
\end{equation}
where $\omega_0$ is the resonance frequency, $\gamma$ represents damping and losses, and $A$ reflects how strongly the side cavities couple to the central chamber. As the incident sound frequency approaches $\omega_0$, the amplification increases rapidly, producing large phase delays and strong acoustic perturbations.

The resonance frequency $\omega_0$ is mainly determined by the effective size $L$ of the metamaterial along the sound propagation direction:
\begin{equation}
\omega_0 \approx \frac{c_{\mathrm{eff}}}{L},
\end{equation}
where $c_{\mathrm{eff}}$ is the effective sound speed within the resonant structure. Since $c_{\mathrm{eff}}$ depends primarily on material properties, it remains approximately constant within the design band. This relationship implies a simple and intuitive rule: larger structures resonate at lower frequencies, while smaller structures resonate at higher frequencies.

We validate this size-frequency relationship using numerical simulations in \textsc{COMSOL Multiphysics}~\cite{comsol}, a finite-element simulator for modeling acoustic wave propagation, resonance, and interference in complex three-dimensional geometries. As shown in Fig.~\ref{structure2}, when the geometric parameters are set to \textbf{$d{=}\mathrm{19.5\,mm}$, $h{=}\mathrm{21\,mm}$, $t{=}\mathrm{1.95\,mm}$, $s{=}\mathrm{49.5\,mm}$}, corresponding to $L \approx \mathrm{779\,mm^2}$, the structure exhibits a resonance centered around 500\,Hz with an interference gain of up to $73\times$. This produces a strong and directional perturbation of voiceprint features in the target band.

We note that varying the metamaterial thickness $z$ has little effect on either the resonance frequency or the interference strength. This is because thickness mainly provides mechanical support and does not significantly alter the effective cavity dimensions or acoustic boundary conditions of the dominant resonance mode. As a result, the design maintains stable acoustic performance while allowing flexibility in thickness to accommodate microphones of different sizes, improving practical deployability.

\subsection{Dynamically Stable Metamaterial Layout}\label{sec:layout}

The band-targeted perturbation described in Sec.~\ref{sec:band_design} is effective when the speaker is stationary and facing the microphone. In practice, speech is often dynamic: speakers naturally turn their heads and adjust their posture, changing the incidence angle and the propagation path of the sound reaching the microphone~\cite{head1,head2}. For direction-sensitive acoustic structures, even small angular deviations can significantly reduce interference strength, leading to inconsistent anonymization results. Such angle sensitivity creates a risk of voiceprint leakage, as identity-bearing features may escape perturbation at certain orientations. Robust voiceprint protection, therefore, requires maintaining strong interference across a wide range of speaking angles, rather than only in a fixed, ideal orientation. 

\SystemName addresses this by designing a dynamically stable metamaterial layout that maintains strong perturbation over common speaking angles. By modeling realistic speaker motion and microphone geometries, we identify a multi-unit arrangement whose combined response provides consistent anonymization under natural user movement, without active sensing or control.

\subsubsection{Anonymization acoustic field model}
A natural way to improve angular coverage is to deploy multiple metamaterial units. However, if the number of units or their orientations are poorly chosen, the sound fields generated by different units can interfere with each other, weakening the overall perturbation effect~\cite{interference1,interference2}. To guide the design, we model and evaluate multi-unit field superposition under dynamic usage through numerical simulation. 

Like Sec.~\ref{sec:band_design}, we build an anonymization acoustic field model in \textsc{COMSOL} to simulate sound pressure distributions under user motion and quantify how the number of metamaterial units and their spatial orientations affect interference strength. Fig.~\ref{model11} illustrates our acoustic field model. To do so, we first define the trajectory of the mouth sound source. Based on prior studies~\cite{headmove1,headmove2}, the mouth orientation varies within $\theta \in [-180^\circ, 180^\circ]$ around the head center, with radius $r_1 \approx 10$\,cm (head-center-to-mouth distance). We thus model the sound source position $(x_{0,\theta}, y_{0,\theta}, z_{0,\theta})$ as:
\vspace{-1mm}
\begin{equation}
\small
\left\{
\begin{aligned}
x_{0,\theta} &= x_0 + r_1 \sin\!\left(\frac{\theta \pi}{180}\right), \\
y_{0,\theta} &= y_0 + r_1 \left(1 - \cos\!\left(\frac{\theta \pi}{180}\right)\right), \\
z_{0,\theta} &= z_0.
\end{aligned}
\right.
\end{equation}
where $(x_0, y_0, z_0)$ is the initial sound source position. We ignore small vertical ($z$-axis) motion because simulations show it has negligible impact on the sound pressure distribution.

Next, we model microphone placement. Because microphone position is not fixed across real-life scenarios, we consider two common portable microphones: a gooseneck microphone and a handheld microphone, as illustrated in Fig.~\ref{model11}. The gooseneck microphone is fixed on the table (it does not rotate with the speaker), but its height can be adjusted to improve capture. To avoid clipping, we set the initial mouth-to-microphone distance to $D \in [10,30]$\,cm~\cite{sounds1}, and the height adjustment range to $H \in [0,30]$\,cm~\cite{gooseee}. 

The gooseneck microphone position $(x_{1,\theta}, y_{1,\theta}, z_{1,\theta})$ is:
\begin{equation}
\small
\left\{
\begin{aligned}
x_{1,\theta} &= x_0, \\
y_{1,\theta} &= y_0 - \sqrt{D^2/2}, \\
z_{1,\theta} &= z_0 - \sqrt{D^2/2} - H,
\end{aligned}
\right.
\end{equation}
where we align $x_{1,\theta}$ with $x_0$ to reflect the common setup in which the speaker faces the microphone.

In contrast, a handheld microphone typically moves with the user and remains approximately aligned with the mouth. Therefore, we model the handheld microphone position $(x_{2,\theta}, y_{2,\theta}, z_{2,\theta})$ as:
\begin{equation}
\small
\left\{
\begin{aligned}
x_{2,\theta} &= x_{1,\theta} + \left(\sqrt{D^2/2} + r_1\right) \sin\!\left(\frac{\theta \pi}{180}\right), \\
y_{2,\theta} &= y_{1,\theta} + \left(\sqrt{D^2/2} + r_1\right) \left(1 - \cos\!\left(\frac{\theta \pi}{180}\right)\right), \\
z_{2,\theta} &= z_0 - \sqrt{D^2/2}.
\end{aligned}
\right.
\end{equation}

By simulating the moving sound source and the trajectories of both microphone types in \textsc{COMSOL}, we obtain sound pressure distributions across angles and positions. This provides a quantitative basis for evaluating multi-unit metamaterial layouts and selecting orientations that remain effective under user motion. Our method is also applicable to other types of microphones, such as those on mobile devices. During mobile calls, users typically hold the microphone near their mouth and move it along with their speech, resulting in a trajectory similar to that of a handheld microphone. When the device is fixed in front of the face for calls (e.g., during video conferences with the camera on), its trajectory can be treated as that of a gooseneck microphone.

\begin{algorithm2e}[!t]
\footnotesize
\SetAlgoLined
\caption{\textit{Dynamically stable anonymization metamaterial layout}}
\label{alg:metanum}
\SetKwProg{Fn}{Function}{:}{}
\Fn{\textsc{DesignLayout}$(\cdot)$}{
$Angle_{M_1} \gets 0^\circ$ \tcp*[r]{Ensure anonymization when facing the microphone}\label{alg:1}

\ForEach{$Angle_{User} \in [-90^\circ, 90^\circ]$}{\label{alg:2}
    \ForEach{$Angle \in [-180^\circ, 180^\circ]$}{
        $Angle_{M_2} \gets Angle$; compute interference gain $I_{M_2}(Angle_{User},Angle_{M_2})$ \label{alg:3}
    }\label{alg:4}
}
$Angle_{M_2}^{best} \gets \arg\max_{Angle_{M_2}} I_{M_2}$ \label{alg:6}

\ForEach{$Angle_{User} \in [-90^\circ, 90^\circ]$}{\label{alg:7}
    \ForEach{$Angle \in [-180^\circ, 180^\circ]$}{
        $Angle_{M_3} \gets Angle$; compute interference gain $I_{M_3}(Angle_{User},Angle_{M_3})$ \label{alg:8}
    }\label{alg:9}
}
$Angle_{M_3}^{best} \gets \arg\max_{Angle_{M_3}} I_{M_3}$ \label{alg:10}

\Return{$Angle_{M_1}, Angle_{M_2}^{best}, Angle_{M_3}^{best}$}
}
\end{algorithm2e}

\subsubsection{Dynamic anonymization}

\begin{figure*}[t!]
  \begin{minipage}[t]{0.483\linewidth}
 \centering
\subfloat[]{
		\includegraphics[scale=0.188]{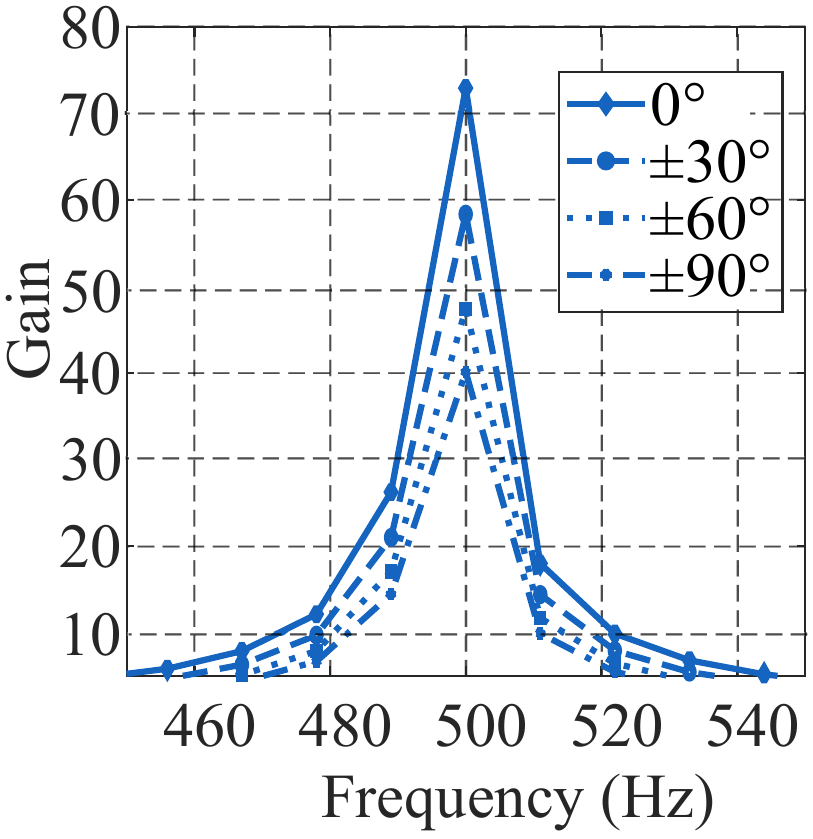}
        \label{aa}}
        \hfill
\subfloat[]{
		\includegraphics[scale=0.188]{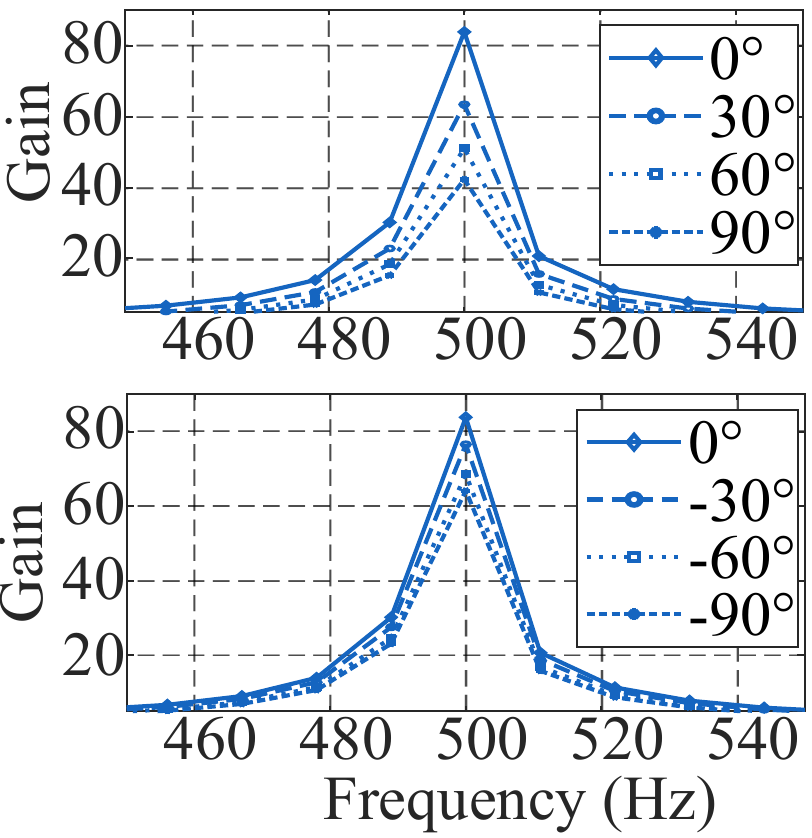}
        \label{bb}}
        \hfill
\subfloat[]{
		\includegraphics[scale=0.188]{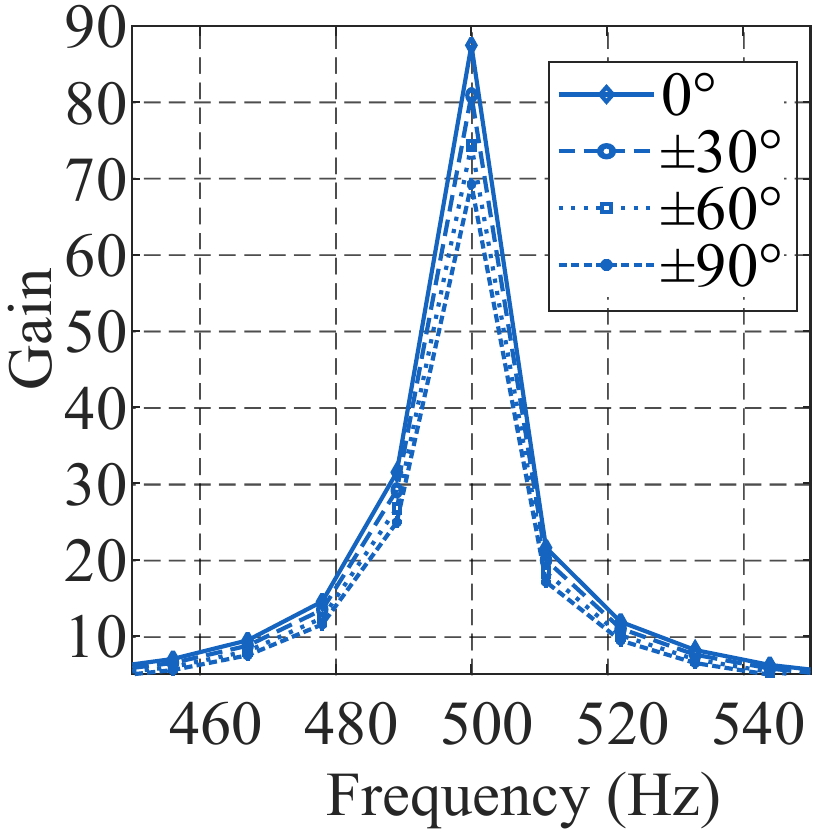}
        \label{cc}}
\caption{(a) Gains at -90°to 90° with M1 only. (b) Top: Gains at 0°to 90° (M1, M2); Bottom: Gains at -90°to 0° (M1, M2). (c) Gains at -90°to 90° with M1, M2, and M3.}
    \end{minipage}  
    \hfill
    \begin{minipage}[t]{0.48\linewidth}
 \centering
\subfloat[]{
    \includegraphics[scale=0.066]{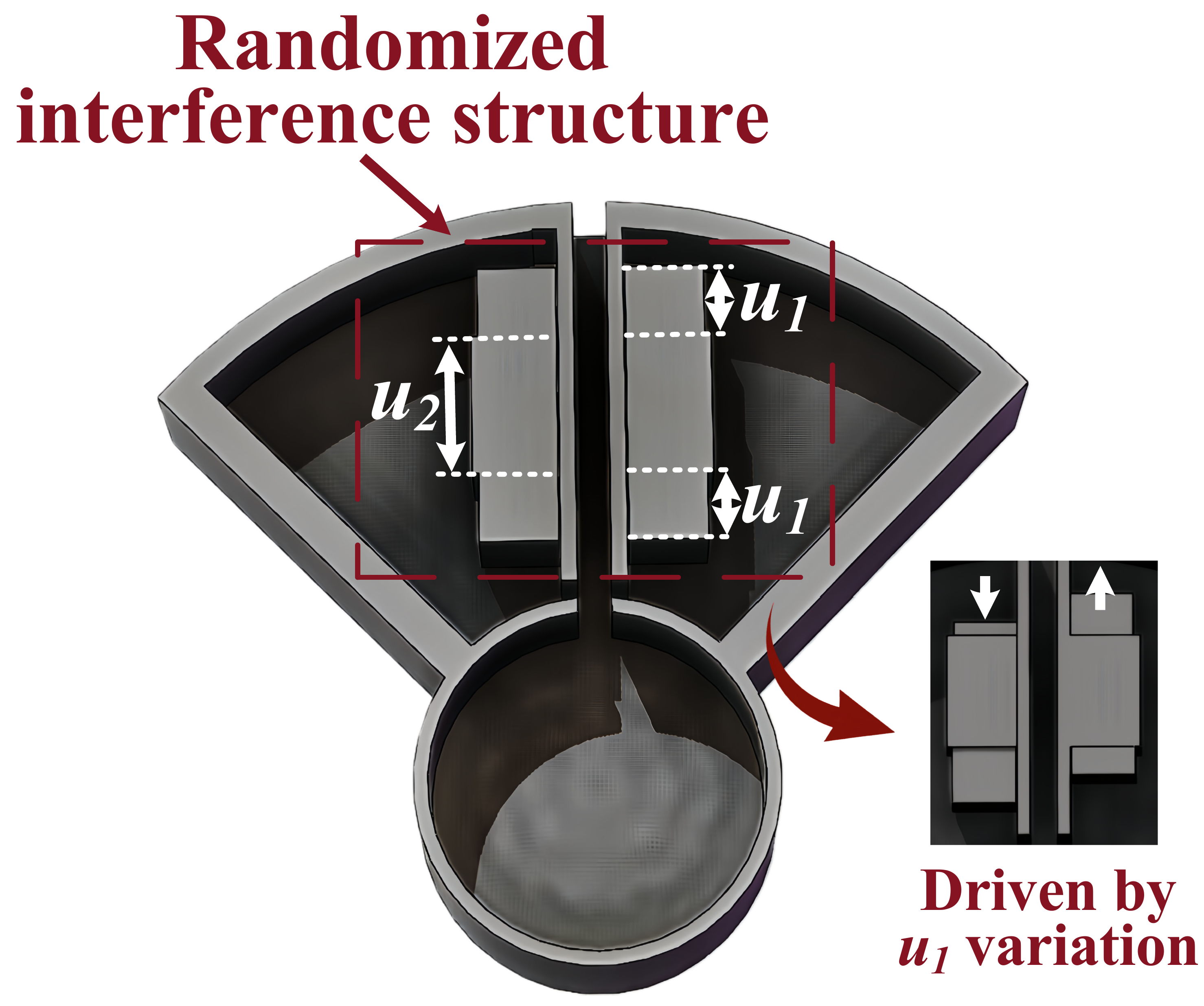}
    \label{C30}}
\hfill
\subfloat[]{
    \includegraphics[scale=0.066]{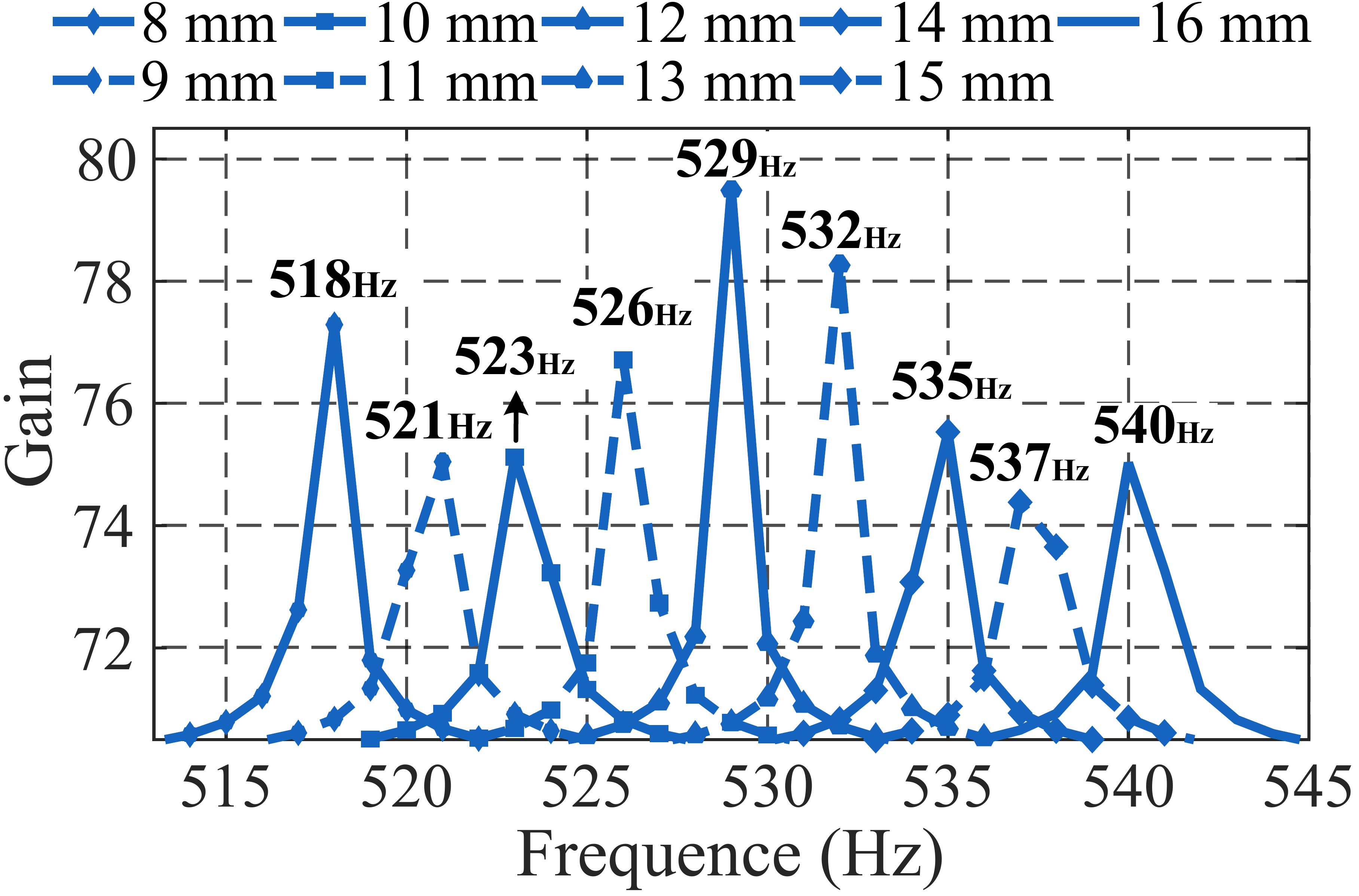}
    \label{C31}}
\caption{(a) Randomized interference metamaterial and (b) its interference performance.}
\label{fig:randomized}
    \end{minipage}  
\end{figure*}

To determine the minimum number of metamaterial units and their orientations, we integrate candidate layouts with the anonymization acoustic field model and evaluate interference gain over the user’s angular range. We start with a single unit $M_1$ oriented at $0^\circ$ (line~\ref{alg:1}), facing the user. This yields strong interference at $0^\circ$, but the gain decreases as the speaking angle deviates, reaching roughly half of the peak at $\pm 90^\circ$ (Fig.~\ref{aa}).


To improve coverage, we add a second unit $M_2$ and scan its orientation over $[-180^\circ, 180^\circ]$ (lines~\ref{alg:2}--\ref{alg:4}). The best orientation, $Angle_{M_2}^{best}=-120^\circ$ (line~\ref{alg:6}), maximizes interference over the left-side range $[-90^\circ,0^\circ]$ and substantially outperforms $M_1$ alone. However, $M_1{+}M_2$ still leaves the $[0^\circ,90^\circ]$ range with relatively low gain (Fig.~\ref{bb}).

We therefore add a third unit $M_3$ and repeat the same search (lines~\ref{alg:7}--\ref{alg:9}). The optimal configuration places $M_3$ symmetrically with respect to $M_2$, i.e., $Angle_{M_3}^{best}=120^\circ$ (line~\ref{alg:10}). This symmetry matches the user’s left--right motion range and yields strong interference across the full angular span $[-90^\circ,90^\circ]$ (Fig.~\ref{cc}).

We apply the same procedure to the handheld microphone case. Because the handheld microphone co-moves with the mouth source and remains approximately aligned, the resulting interference gains are close to the gooseneck case at $0^\circ$ (Fig.~\ref{cc}). Based on these results, we adopt the three-unit symmetric layout as our dynamically stable metamaterial configuration (Fig.~\ref{model22}).

\subsection{Passive Randomization}\label{sec:ril}

While our designs so far achieve stable anonymization across different scenarios and user movements, a fixed interference pattern compromises security guarantees. Acoustic metamaterials are inherently passive, and their geometric configurations are largely fixed after fabrication. As a result, their acoustic responses remain stable and predictable over time, which may expose the system to observation-based or modeling-based reverse-engineering attacks \cite{attacks2}.

To further enhance robustness without sacrificing passivity, we introduce controlled randomness into the acoustic response itself. The key idea is to generate dynamic and unpredictable interference patterns using purely, power-free physical mechanisms that respond to natural user motion. This enables continuous randomization of the interference behavior while preserving usability and deployment simplicity.

\subsubsection{Randomized interference}


To this end, our design introduces dynamic perturbations to the interference curve while remaining fully passive. As shown in Fig.~\ref{C30}, the design introduces a slidable block with adjustable length $u$ inside the internal structure of a metamaterial unit. The block consists of a telescopic segment of length $u_1$ and a hollow outer segment of length $u_2$, into which $u_1$ can extend. Variations in $u$ change the occupied cavity volume, thereby modulating the effective acoustic boundary and the resonance conditions.

This mechanism can be modeled as a modulation of the effective spatial size $L$ of the metamaterial (Sec.~\ref{sec:band_design}):
\vspace{-2mm}
\begin{equation}
L(u) = L_0 - \gamma u,
\end{equation}
where $L_0$ is the nominal effective size and $\gamma$ depends on the geometry of the sliding block. Substituting this relation into the resonance condition yields
\begin{equation}
\omega_0(u) \approx \frac{c_{\mathrm{eff}}}{L_0 - \gamma u}.
\end{equation}
As $u$ increases, $L$ decreases, shifting the resonance frequency toward higher values. Importantly, $u$ varies naturally as the block moves along its guide in response to small user motions, inducing random yet continuous frequency fluctuations. This produces time-varying interference patterns that are difficult to observe or model, significantly increasing resistance to reverse engineering.

Although the width and height of the sliding block also influence $L$, their adjustable ranges are limited by the internal cavity dimensions and have effects similar to $u$. We therefore use $u$ as the primary source of randomization, fixing the block width and height at approximately 5 mm. This configuration allows sufficient movement without mechanical interference while providing effective resonance modulation.

To avoid excessive frequency drift that could harm anonymization, the adjustment range of $u$ is constrained. The total block length is limited to 16 mm, with the telescopic segment $u_1$ varying within 4\,mm and the hollow segment $u_2$ fixed at 8mm. Under these constraints, the interference center frequency remains within the target 50Hz band, and the interference amplitude varies only marginally.

Again, we validate our design using \textsc{COMSOL Multiphysics}. As depicted in Fig.~\ref{C31}, changes in $u$ shift the interference center frequency as expected, while the interference gain remains consistently above $72\times$. This confirms that passive randomization effectively enhances security without degrading voiceprint perturbation performance. Later in our evaluation, we show that \SystemName performs well in real-life settings. 




\begin{figure*}[t!]
  \begin{minipage}[t]{0.48\linewidth}
    \centering
    \includegraphics[width=0.92\linewidth]{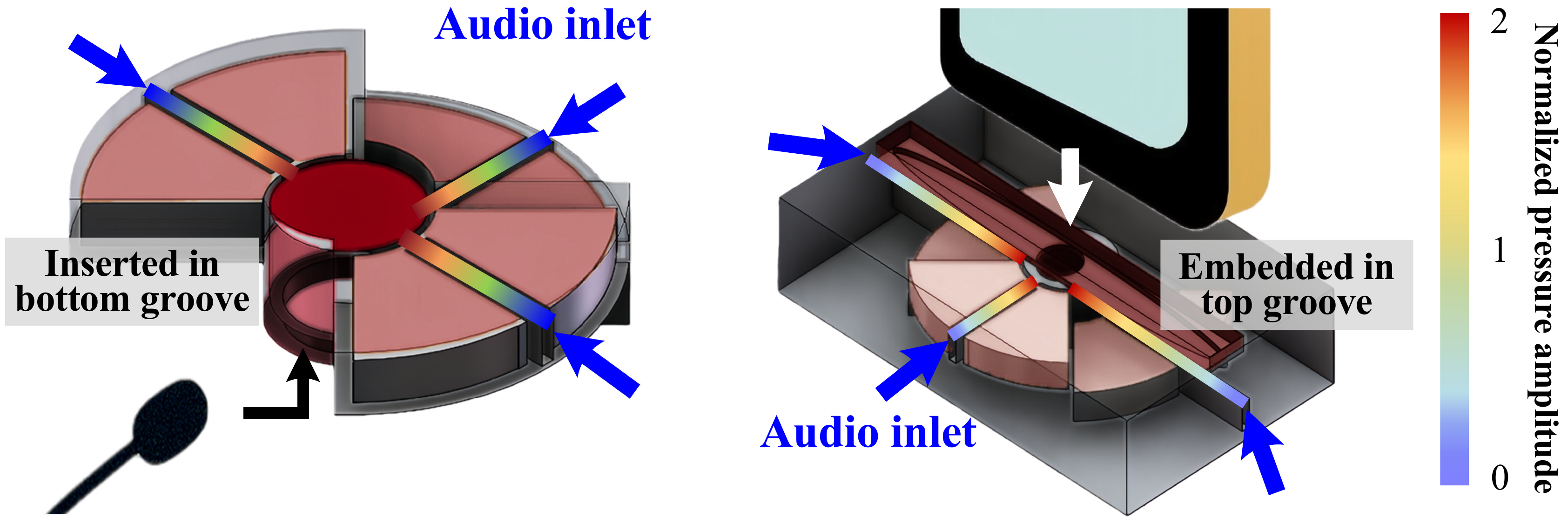}
    \caption{Functional variants of system: compatible with conventional microphones (left) and mobile devices (right).}
    \label{fig:prototype2}
    \end{minipage}  
    \hfill
    \begin{minipage}[t]{0.48\linewidth}
    \centering
    \includegraphics[width=0.99\linewidth]{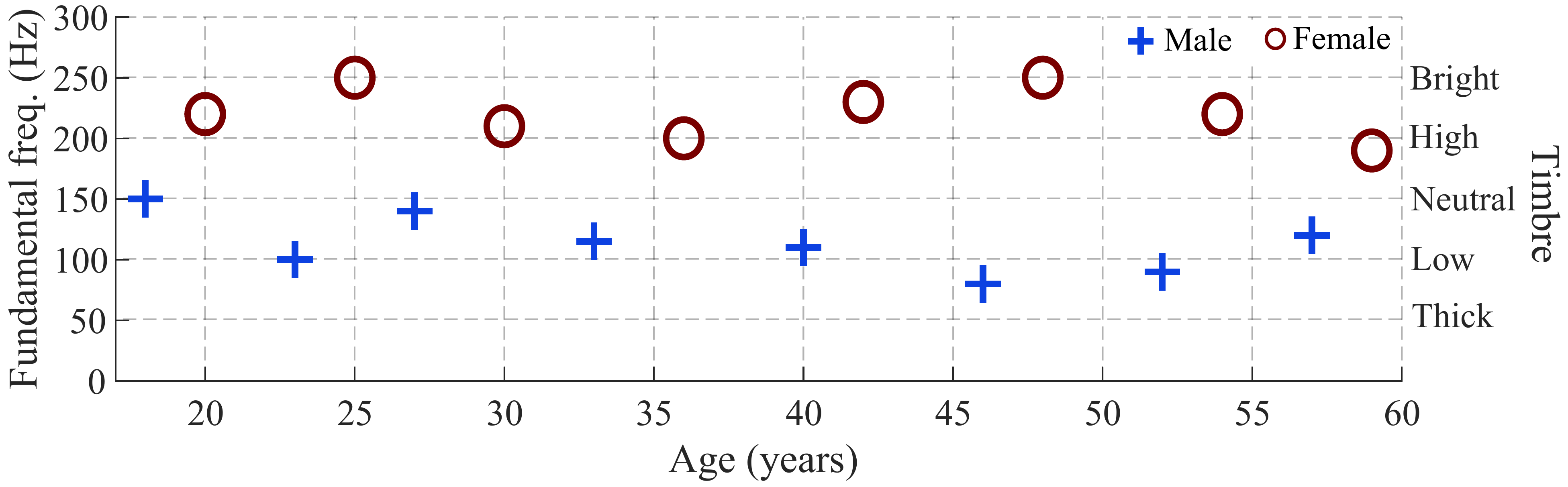}
    \caption{Participant characteristics in the user study: gender, age, fundamental frequency (pitch) and timbre variances.}
    \label{vo}
    \end{minipage}  
\end{figure*}

Considering these contributions and the structural differences between traditional and mobile-device microphones, we designed two enclosure variants of \SystemName, as shown in Fig.~\ref{fig:prototype2}. For devices of different sizes, only minor adjustments to the groove dimensions are required. As these changes do not alter the core metamaterial structure, the interference performance remains unaffected.

\section{Experimental Setup}
All our experiments were conducted under approval from the Institutional Review Board (IRB). The research equipment was self-funded, and participants voluntarily joined with informed consent. No sensitive or personally identifying data was collected or stored during the study, ensuring compliance with ethical standards.  

\begin{table}[t!]
    \scriptsize
    \caption{ASV models and microphone devices used in the evaluation (including 5 models and 8 devices).}
    \label{modelsdevices}
    \vspace{1mm}
    \centering
    \setlength{\tabcolsep}{12pt} 
    \begin{tabular} {llllll}
    \toprule
    \makecell[l]{\textbf{Model / Device}} &  \textbf{Category} & \textbf{Source / Manuf.}\\
    \midrule
    \rowcolor{gray!20} iFlytek~\cite{iFlytekASV} & Commercial & iFlytek \\
    \rowcolor{gray!20} ECAPA-TDNN~\cite{ECAPA-TDNN}  & DNN-based &  SpeechBrain\\
    \rowcolor{gray!20} X-vector~\cite{Xvectors} & DNN-based & SpeechBrain \\
    \rowcolor{gray!20} GMM-UBM~\cite{GMMUBM} &  Statistical  &  Sidekit \\  
    \rowcolor{gray!20} ivector-PLDA~\cite{ivectorPLDA} & Statistical &  Kaldi\\
    
    Shure SV200~\cite{Shuresv200} &  Handheld mic  &  Shure \\     
   Behringer TA5212~\cite{BehringerTA5212} & Gooseneck mic  & Behringer \\     
   Audio‑Technica AT9930~\cite{Audio-TechnicaAT9930} & Gooseneck mic & Audio‑Technica\\
   sE Electronics V7~\cite{sEElectronicsV7} & Handheld mic & sE Electronics\\
    \rowcolor{gray!20}iPhone 16 Pro Max~\cite{Apple} & Mobile device & Apple\\
   \rowcolor{gray!20}Pixel 8 Pro~\cite{Googlep} & Mobile device & Google\\
   \rowcolor{gray!20}Mate 60 Pro~\cite{Huawei} & Mobile device & Huawei\\
      \rowcolor{gray!20}Galaxy S24~\cite{samsungS24} & Mobile device & Samsung\\
    \bottomrule
    \end{tabular}
\end{table}

\subsection{Test Targets}
As summarized in Table~\ref{modelsdevices}, we evaluate \SystemName against five mainstream automatic speaker verification (ASV) systems to assess its effectiveness in preventing voiceprint abuse. To examine generalizability across hardware, we conduct experiments using microphones from eight different manufacturers. As shown in Fig.~\ref{vo}, we recruit 16 volunteers with balanced gender representation, diverse age groups, and varied timbral characteristics to complete the reading tasks in our experimental corpus. These recordings are used to evaluate anonymization performance under different speaker conditions (Sec.~\ref{sec:bp}). In addition, we recruit 50 gender-balanced volunteers to participate in a subjective listening study, assessing the intelligibility and perceived quality of anonymized speech (Sec.~\ref{sec:epq}).


\subsection{Evaluation Metrics}
Following common practice in prior work~\cite{vcloak,micpro,vsmask}, we evaluate \SystemName using four complementary metrics: \emph{Miss-Match Rate (MMR)}, \emph{Word Accuracy (WA)}, \emph{Mean Opinion Score (MOS)}, and \emph{Real-time Coefficient (RTC)}, as summarized in Table~\ref{Evaluating metrics}. Together, these metrics capture the core goals of voiceprint anonymization: identity protection, speech intelligibility, perceptual quality, and practical deployability.

\emph{MMR} quantifies the effectiveness of voiceprint protection by measuring how often anonymized speech fails speaker matching. For each device-condition pair, we perform 30 trials; anonymization is considered successful if the voiceprint similarity between the anonymized and original audio falls below a threshold of 0.25, following prior work~\cite{vsmask}. This metric directly reflects resistance to speaker recognition attacks.

\emph{WA} evaluates whether speech content remains intelligible after anonymization. We compute WA using Google Speech Recognition by measuring the proportion of correctly recognized words in the anonymized audio~\cite{vcloak,micpro,vsmask}. Since ASR systems are designed to recover linguistic content under noise and distortion, WA serves as a conservative proxy for human speech intelligibility.

\emph{MOS} captures subjective perceptual quality from the listener's perspective. We collect MOS scores from 10 gender-balanced volunteers, who rate the perceived quality of anonymized speech on a 5-point scale (1 = Bad, 5 = Excellent), following established evaluation protocols~\cite{vcloak,micpro}. This metric reflects how natural and usable the anonymized speech sounds to humans.

Finally, \emph{RTC} measures system efficiency and suitability for real-time use. It is defined as
\vspace{-2mm}
\[
\text{RTC} = \frac{T_{\text{cvt}}}{T_{\text{audio}}},
\]
where \(T_{\text{audio}}\) is the audio duration and \(T_{\text{cvt}}\) is the anonymization time. Lower RTC values indicate higher efficiency~\cite{vcloak,micpro}. This metric ensures that anonymization can be applied at capture time without introducing prohibitive latency.

 \begin{table}[t!]
    \scriptsize
    \caption{Evaluating metrics (covering four common metrics).}
    \label{Evaluating metrics}
    \vspace{1mm}
    \centering
     \setlength{\tabcolsep}{10pt}
    \begin{tabular} {lp{6.3cm}}
    \toprule
    \multicolumn{1}{l}{\textbf{Metrics}} & \multicolumn{1}{l}{\textbf{Description}}\\
    \midrule
\rowcolor{gray!20} MMR & Miss-Match Rate (MMR) is the proportion of anonymized audios with a voice similarity score below 0.25 out of 30 tests~\cite{vcloak,micpro,vsmask}. \\ 

WA & Word Accuracy (WA) quantifies the proportion of correctly recognized words in anonymized audio via ASR~\cite{vcloak,micpro,vsmask}.  \\ 

\rowcolor{gray!20} MOS & Mean Opinion Score (MOS) assesses the similarity of anonymized vs. original audio through human ratings~\cite{vcloak,micpro}.\\

\makecell[l]{RTC} & Real-time Coefficient (RTC) is used to evaluate the processing efficiency of \SystemName for voice anonymization~\cite{vcloak,micpro}. \\

\bottomrule
\end{tabular}
\end{table}

\begin{table*}
    \scriptsize
    \caption{Test scenarios of our evaluation}
    \label{experimental setting}
    \vspace{1mm}
    \centering
    \renewcommand{\arraystretch}{1.2}
    \begin{tabular} {p{1.6cm}p{1.6cm}p{4.3cm}p{8.9cm}} \toprule
    \multicolumn{1}{l}{\textbf{Objectives}}  & \multicolumn{1}{l}{\textbf{Label}} &\multicolumn{1}{l}{\textbf{Test focus}} & \multicolumn{1}{l}{\textbf{Description}}\\
    \midrule

\rowcolor{gray!20}  & A1 (Sec.~\ref{sec:bp}) & Impact of microphone models & Evaluated the impact of different microphone models on \SystemName.\\

\rowcolor{gray!20} & A2 (Sec.~\ref{sec:bp}) & Impact of different speakers & Evaluated the impact of speakers with different genders and ages on \SystemName.\\

\rowcolor{gray!20} & A3 (Sec.~\ref{sec:bp}) & Impact of speaking volume & Evaluated the impact of different speaking volume levels on \SystemName's performance.\\

\rowcolor{gray!20}& A4 (Sec.\ref{sec:bp}) & Impact of semantic content & The system's robustness was evaluated across semantics.\\

\rowcolor{gray!20}& A5 (Sec.\ref{sec:epq}) & Processing efficiency & Evaluated the anonymization efficiency of the system.\\

\rowcolor{gray!20} \multirowcell{-3.5}{\makecell[l]{Robustness,\\ usability,\\ and efficiency}} & A6 (Sec.~\ref{sec:epq}) & Human subjective auditory & Subjective audibility was evaluated by comparing anonymized audio with the original audio.\\

& B1 (Sec.~\ref{B1}) & Effect on audio accuracy & Evaluated the effect of anonymized audio on ASR accuracy.\\

& B2 (Sec.~\ref{B2}) & Performance in dynamic environments & Evaluated the anonymization performance of microphones at different positions.\\

\multirowcell{-2}{ \makecell[l]{Quantification\\ of contribution}} & B3 (Sec.~\ref{B3}) & Contribution of complex interference design & Evaluated the robustness of \SystemName's complex interference design.\\

\rowcolor{gray!20} & C1 (Sec.~\ref{C1}) & Impact of mobile environments & Evaluated the impact of speaker movement in remote meetings on system performance.\\

\rowcolor{gray!20} & C2 (Sec.~\ref{C2}) & Impact of noise environment & Evaluated the effect of different noise levels on system performance.\\

\rowcolor{gray!20} \multirowcell{-2}{\makecell[l]{Real scenarios'\\ performance }} & C3 (Sec.\ref{C3}) & Impact of wind speed & Impact of different wind speed on the \SystemName.\\

\bottomrule
\end{tabular}
\end{table*}

\subsection{Experiment Design}
\begin{figure}[!t]
\centering
\subfloat[]{
    \includegraphics[scale=0.0482]{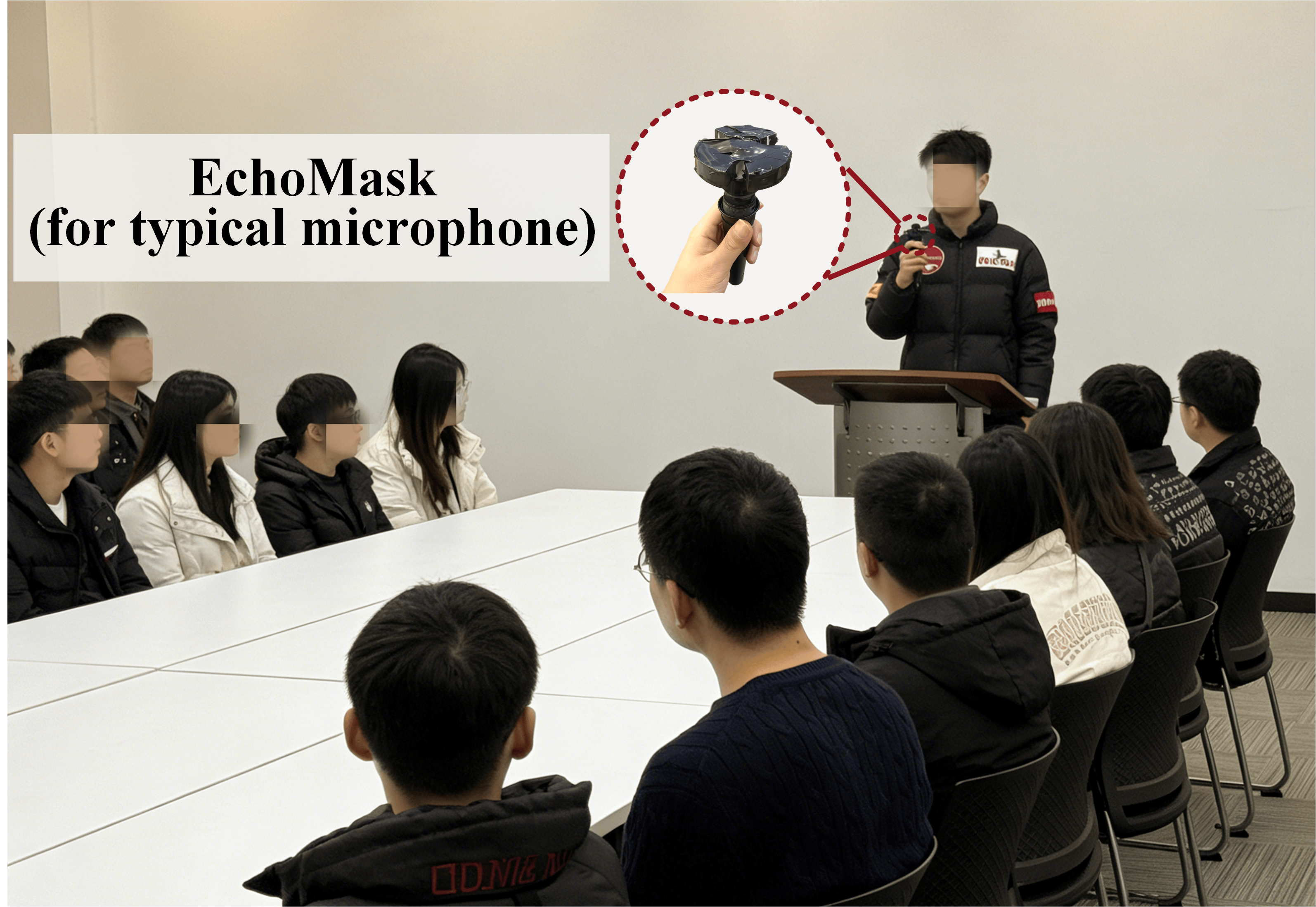}
    \label{S11}}
\hfill
\subfloat[]{
    \includegraphics[scale=0.0482]{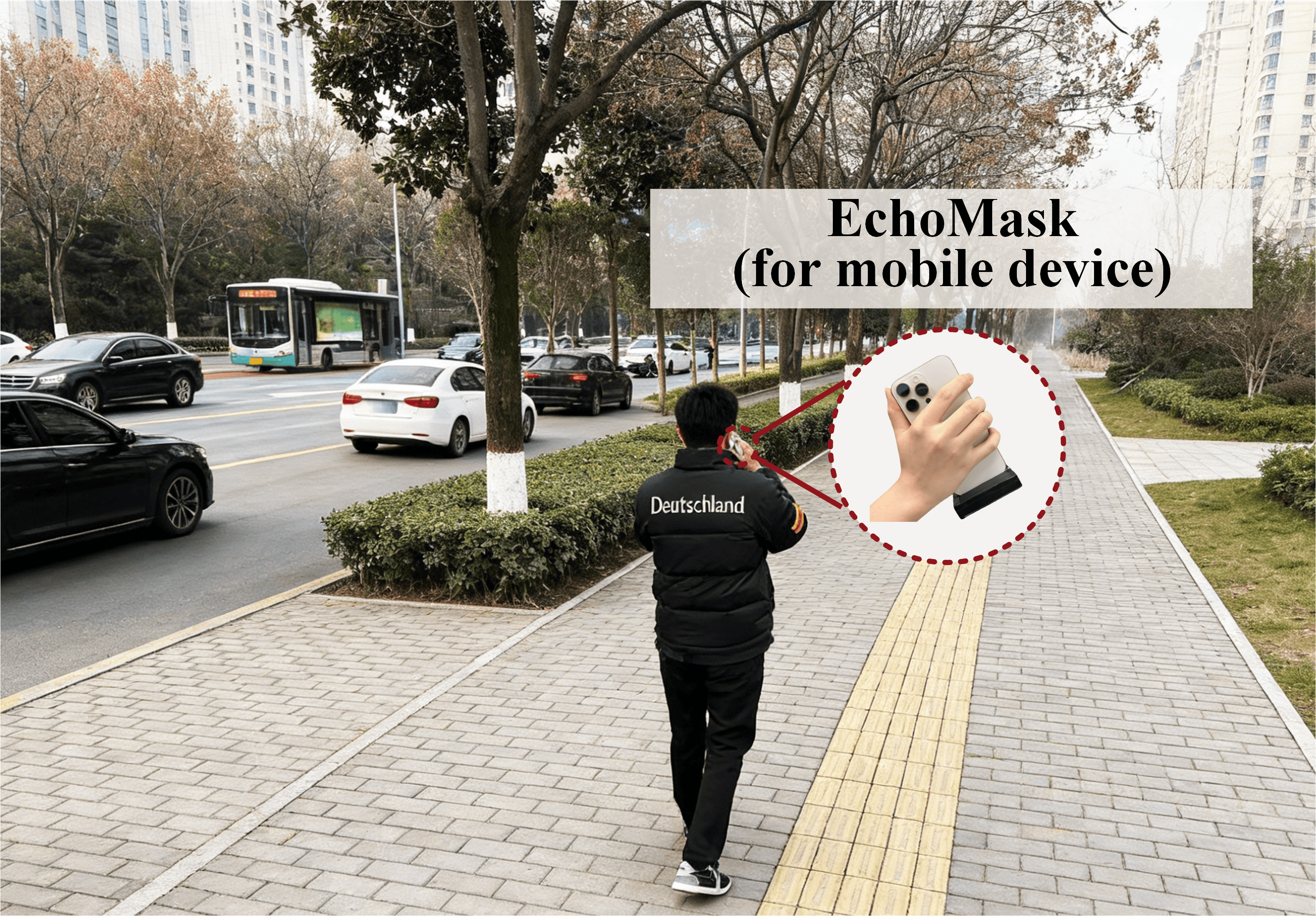}
    \label{S22}}
\caption{Experimental environments: public speech (a) and outdoor mobile phone call (b) scenarios.}
\label{S1122}
\end{figure}

Our experiments aim to simulate realistic real-time call and speech usage scenarios. As shown in Fig.~\ref{S1122}, the experiments were conducted in both an open meeting room and outdoor environments. Specifically, the experiments in Sec.~\ref{sec:rue} and Sec.~\ref{sec:ecs} were carried out in the open meeting room (Fig.~\ref{S11}), while part of the experiments in Sec.~\ref{sec:prw} were performed in outdoor settings (Fig.~\ref{S22}).

In each trial, a volunteer spoke a pre-scripted passage with 650 words (``\textit{Voiceprint anonymization is an important technology…}’’, see Sec.~\ref{OS} for the full text) at a sound pressure level of about 70 dB, closely reflecting continuous speech in real-world usage. The speech lasted about seven minutes. \SystemName was mounted on the devices listed in Table~\ref{modelsdevices}, and the final results were obtained by averaging the outcomes across trials.

\section{Experimental Results} \label{chap:6}
We organize our evaluation and discussion around four aspects. \textbf{Sec.~\ref{sec:rue}} presents baseline performance tests to validate system functionality and key components. \textbf{Sec.~\ref{sec:ecs}} evaluates design choices and ablation studies to assess the impact of individual design elements. \textbf{Sec.~\ref{sec:prw}} examines system performance in real-world scenarios, including various types of environmental noise and wind interference. 

\subsection{Robustness, Usability, and Efficiency}\label{sec:rue}

We start by evaluating \SystemName across three complementary dimensions critical to practical deployment: robustness under heterogeneous devices and speech variances, impact on speech usability, and processing efficiency. Unless otherwise stated, all results report the Miss-Match Rate (MMR) averaged across five representative speaker recognition systems as described in Table.~\ref{modelsdevices}. 

\subsubsection{Robustness across hardware and speakers}\label{sec:bp}
\begin{figure}[t!]
    \centering
    \includegraphics[width=1\linewidth]{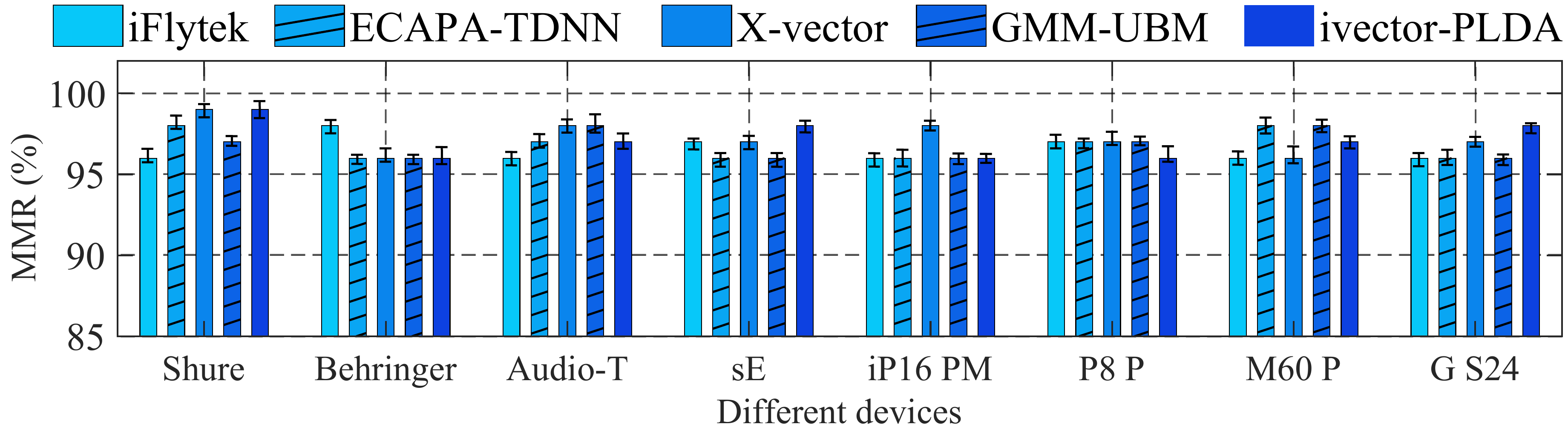}
    \caption{Impact of different devices on anonymization performance (including eight different devices).}
    \label{diffdevices}
\end{figure}

\paragraph{Impact of microphone models.}
As shown in Fig.~\ref{diffdevices}, \SystemName consistently achieves an MMR exceeding 95\% across eight microphone devices and all evaluated speaker recognition frameworks. Performance variation across devices is less than 4\%, indicating that microphone differences have minimal impact on \SystemName's anonymization effectiveness. This robustness is because that metamaterial interference is applied \emph{before} sound waves reach the microphone, making the protection largely independent of downstream capture characteristics.

\begin{figure}[t!]
    \centering
    \includegraphics[width=1\linewidth]{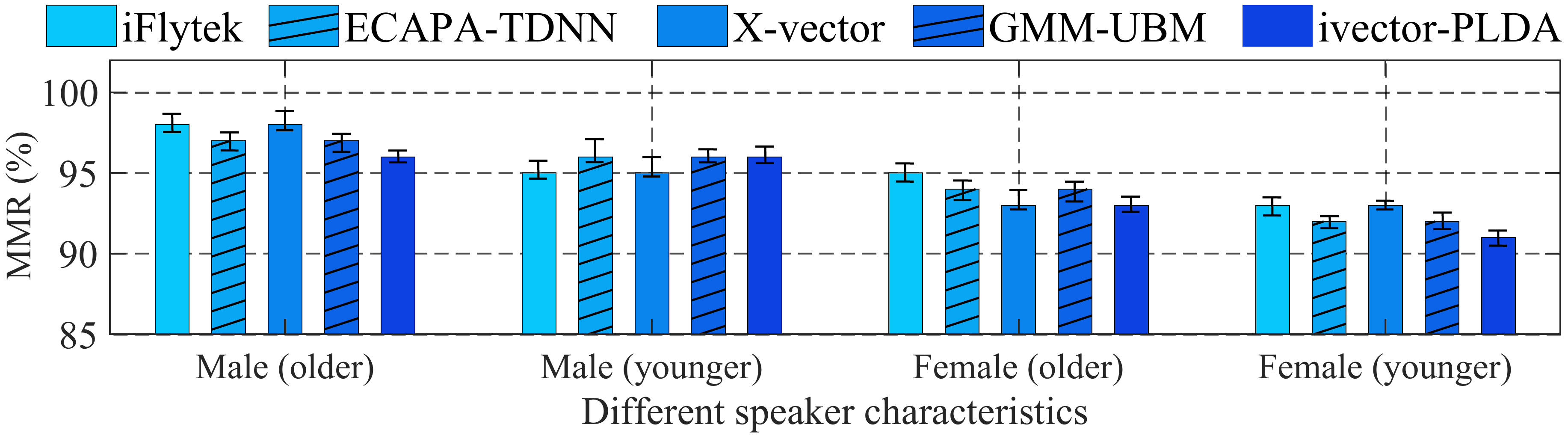}
    \caption{Impact of speaker differences on anonymization performance (younger refers to speakers under 40 and older to speakers 40 and above).}
    \label{diffspeakers}
\end{figure}

\paragraph{Impact of speaker characteristics.}
Fig.~\ref{diffspeakers} shows that \SystemName achieves MMRs above 95\% for male and older speakers, and above 90\% for female and younger speakers. The slightly higher performance for male and older speakers is consistent with their typically lower first formant frequencies, which align more closely with the targeted low-frequency perturbation band. Importantly, anonymization remains effective across all speaker groups, demonstrating strong generality.

\begin{figure*}[t!]
  \begin{minipage}[t]{0.47\linewidth}
    \centering
    \includegraphics[width=1\linewidth]{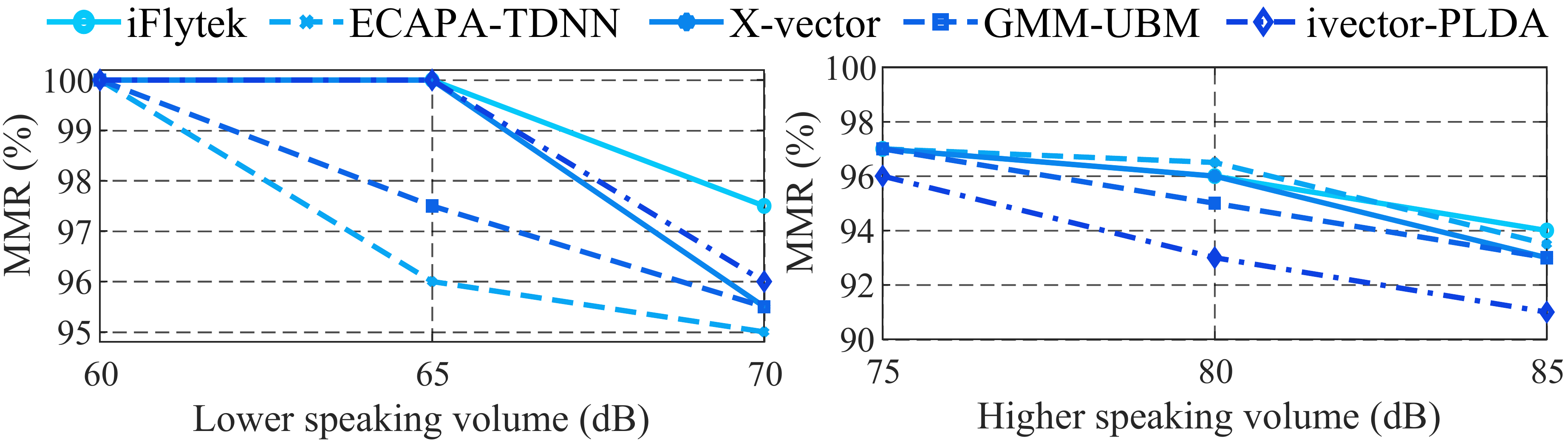}
    \caption{Impact of volume on anonymization performance (covering low to high volume levels from 60 to 85 dB)}
    \label{volume}
    \end{minipage}  
    \hfill
    \begin{minipage}[t]{0.48\linewidth}
    \centering
    \includegraphics[width=1\linewidth]{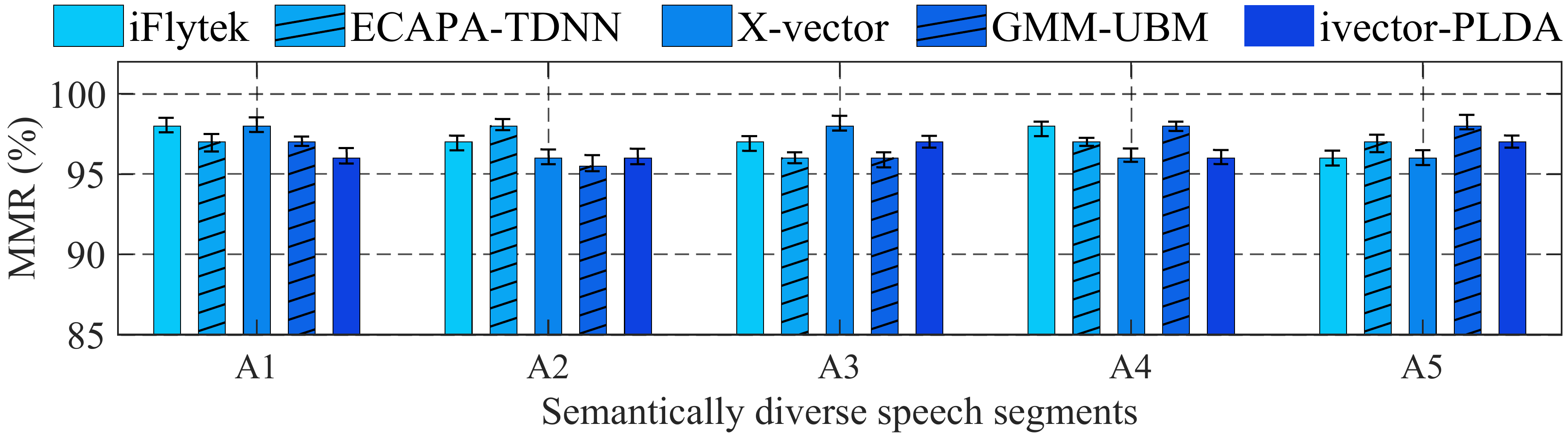}
    \caption{Impact of different semantic content on anonymization performance (including five different audio segments).}
    \label{Semantic}
    \end{minipage}  
\end{figure*}

\paragraph{Impact of speaking volume.}
Speaking volume affects acoustic energy distribution and may alter interference behavior. We therefore evaluate \SystemName across a range of realistic speaking volumes from 60 to 85\,dB.
As shown in Fig.~\ref{volume}, \SystemName maintains an MMR above 90\% across the entire volume range. At lower volumes (60–70\,dB), MMR exceeds 95\%, likely because the interference signal constitutes a larger fraction of the total acoustic energy, strengthening the disruption of speaker-specific features.


\paragraph{Impact of semantic content.}
Different semantic content introduces variation in phonetic structure, prosody, and spectral distribution. To assess sensitivity to content variation, we evaluate \SystemName across five distinct semantic scenarios (see Sec.~\ref{OS}).
Fig.~\ref{Semantic} shows that MMR remains above 95\% across all semantic categories, with differences below 3\%. This indicates that semantic variation has minimal impact on anonymization performance, consistent with our design goal of targeting identity-related acoustic features rather than linguistic content.

\subsubsection{Efficiency and perceptual quality}\label{sec:epq}
\begin{figure}[!t]
\centering
\subfloat[]{
    \includegraphics[scale=0.123]{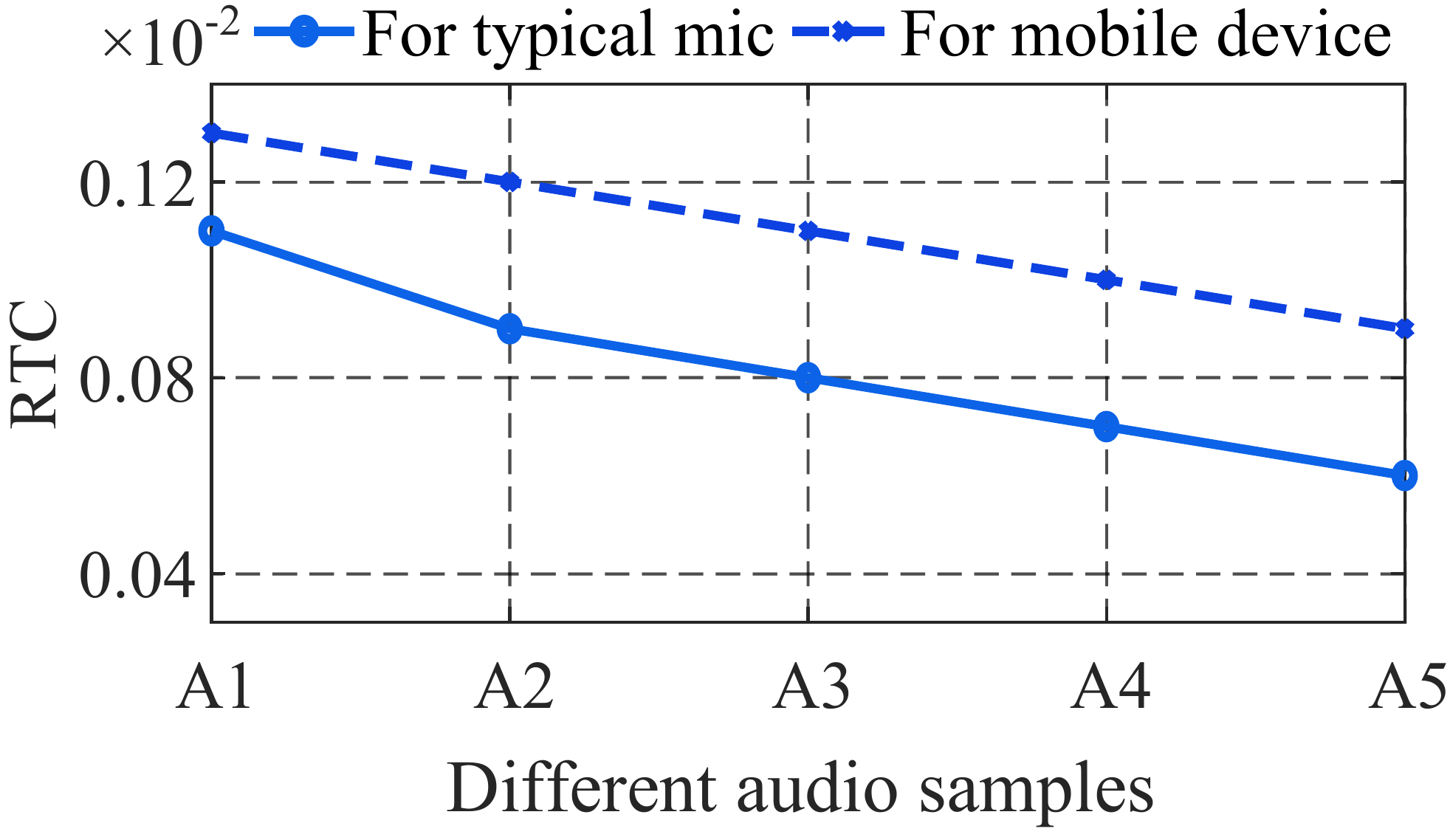}
    \label{RTC}}
\hfill
\subfloat[]{
    \includegraphics[scale=0.123]{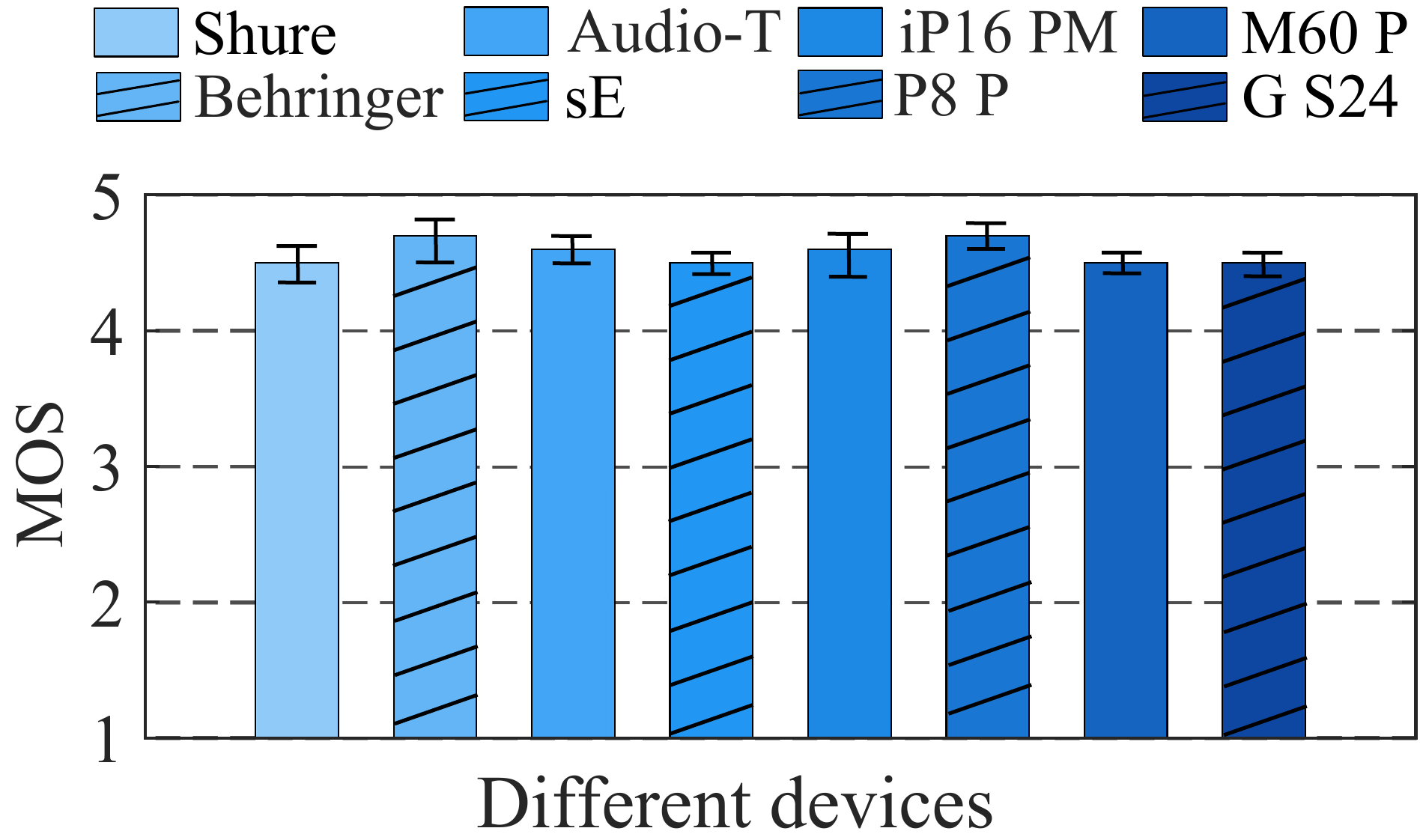}
    \label{MOS}}
\caption{(a) Anonymization efficiency of \SystemName and (b) the subjective auditory of the anonymized audio.}
\label{RTCMOS}
\end{figure}

\paragraph{Processing efficiency.}
Real-time performance is essential for deployment in live settings such as talks, meetings, and online conferences. We evaluate efficiency using the Real-time Coefficient (RTC)~\cite{vcloak,micpro} on audio samples of varying lengths and content.
Because \SystemName operates purely at the physical layer, its delay is dominated by sound propagation through the metamaterial channels. Fig.~\ref{RTC} shows that the mobile-device and conventional-microphone configurations achieve RTC values below 0.0013. Although the mobile configuration exhibits slightly higher delay due to longer internal channels, the overall latency remains negligible.

\paragraph{Human subjective auditory.}
Finally, we evaluate perceived speech quality via a human listening study with 50 gender-balanced volunteers using a 5-point MOS. As shown in Fig.~\ref{MOS}, the average MOS for intelligibility, clarity, and naturalness all exceed 4 across devices, indicating that \SystemName maintains high perceived speech quality despite strong voiceprint disruption. This is mainly because the narrowband, low-frequency perturbation selectively affects speaker identity cues while largely preserving perceptually important speech components.

\subsection{Enhanced Capability of \SystemName\label{sec:ecs}}

\subsubsection{Effect on audio accuracy}\label{B1}
\begin{figure}[t!]
    \centering
    \includegraphics[width=1\linewidth]{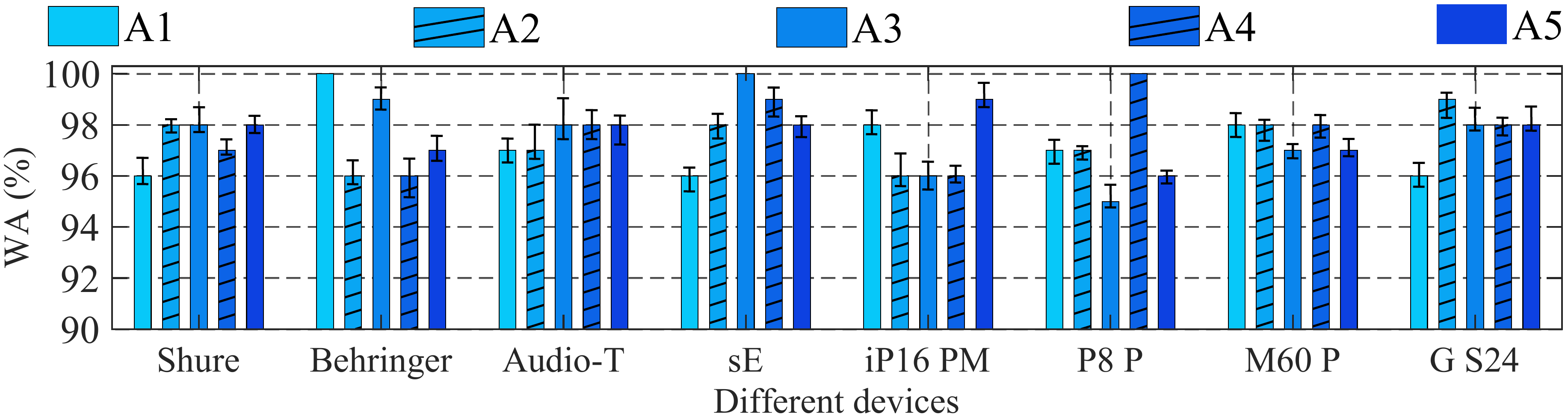}
    \caption{Impact of the system on ASR performance (covering eight different devices and five audio segments)}
    \label{ASRTEST}
\end{figure}
Because the interference band partially overlaps with the speech spectrum, naive anonymization can degrade speech recognition accuracy. To mitigate this, \SystemName adopts a selective interference scheme that targets a narrow set of identity-critical frequencies as described in Sec.~\ref{sec:tlp}, rather than applying broad-spectrum distortion. 
To evaluate the effectiveness of this design, we anonymize speech with different semantic content (Sec.~\ref{OS}) across multiple devices and transcribe the resulting audio using Google Speech-to-Text~\cite{Google}. We then measure speech recognition accuracy on the anonymized recordings. As shown in Fig.~\ref{ASRTEST}, \SystemName consistently achieves over 95\% transcription accuracy across devices and speech contents. This indicates that the proposed interference scheme has only a minor impact on speech recognition, validating its ability to preserve intelligible and usable speech while disrupting voiceprints.

\subsubsection{Performance in dynamic environments}\label{B2}
\begin{figure*}[t!]
  \begin{minipage}[t]{0.47\linewidth}
    \centering
\subfloat[]{
    \includegraphics[scale=0.12]{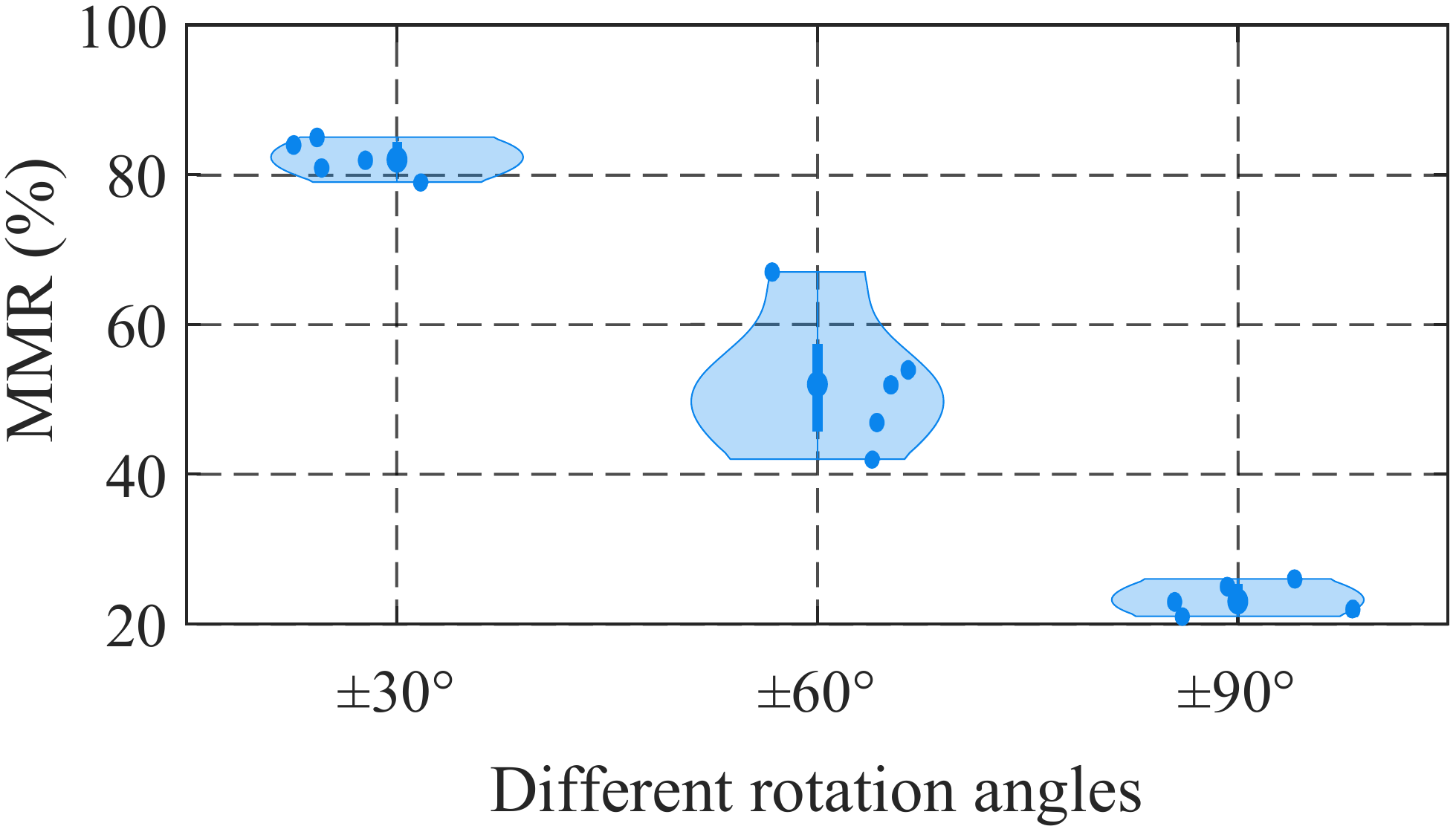}
    \label{ST1}}
\hfill
\subfloat[]{
    \includegraphics[scale=0.12]{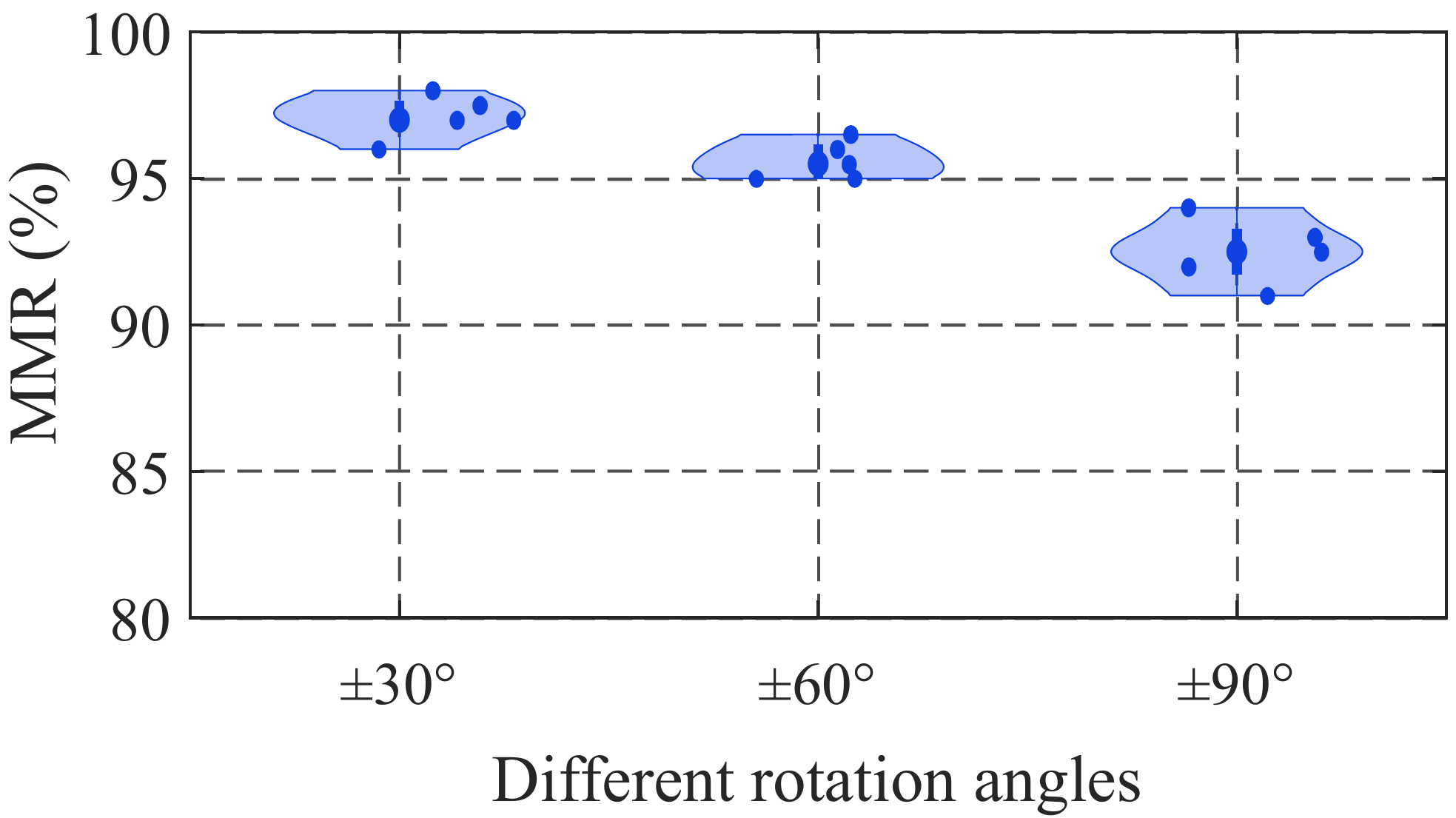}
    \label{ST2}}
\caption{(a) Anonymization performance without the dynamically stable structure, (b) with structure deployed.}
\label{ST}
    \end{minipage}  
    \hfill
    \begin{minipage}[t]{0.48\linewidth}
    \centering
\subfloat[]{
		\includegraphics[scale=0.117]{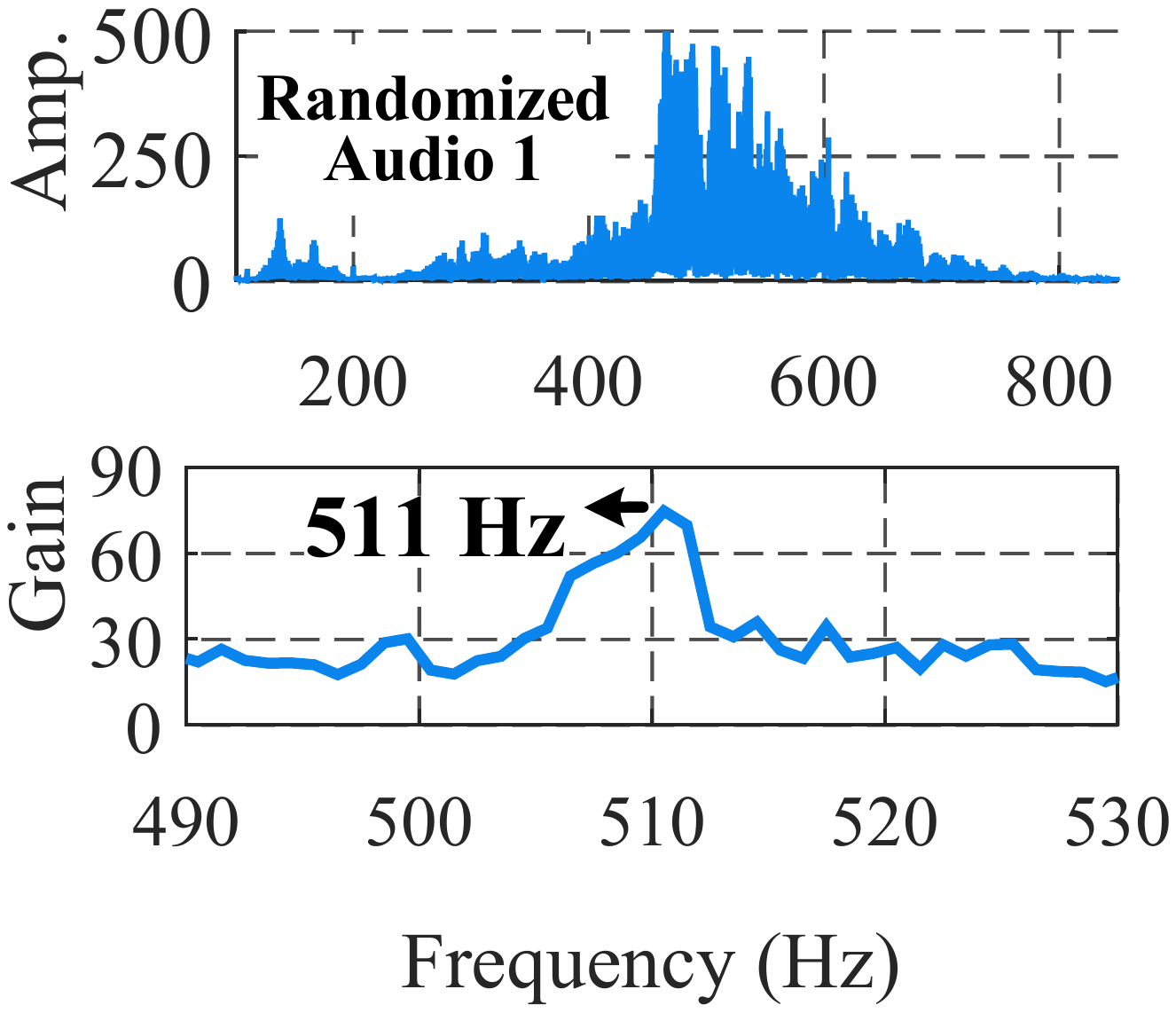}
        \label{R1}}
        \hfill
\subfloat[]{
		\includegraphics[scale=0.117]{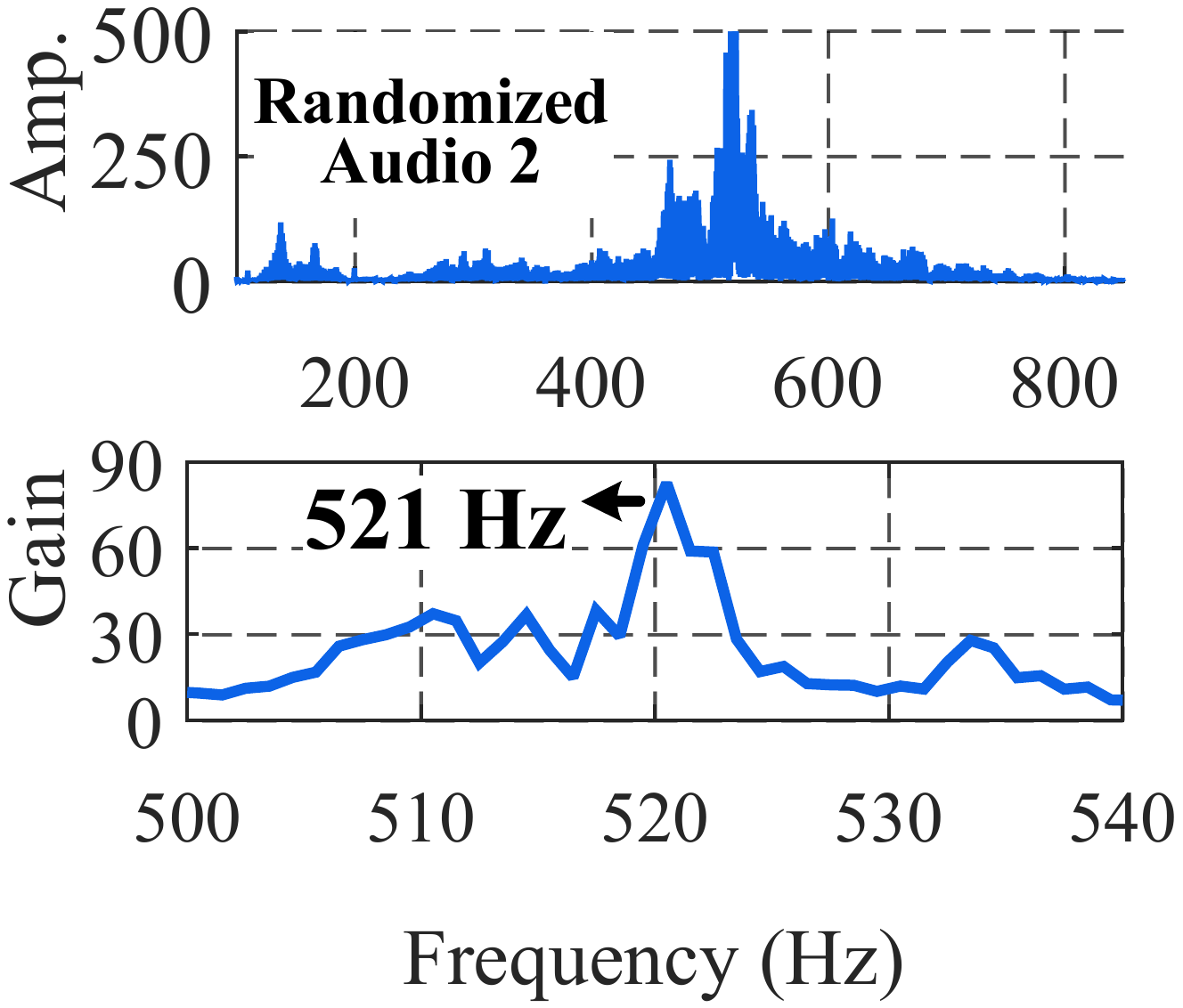}
        \label{R2}}
        \hfill
\subfloat[]{
		\includegraphics[scale=0.117]{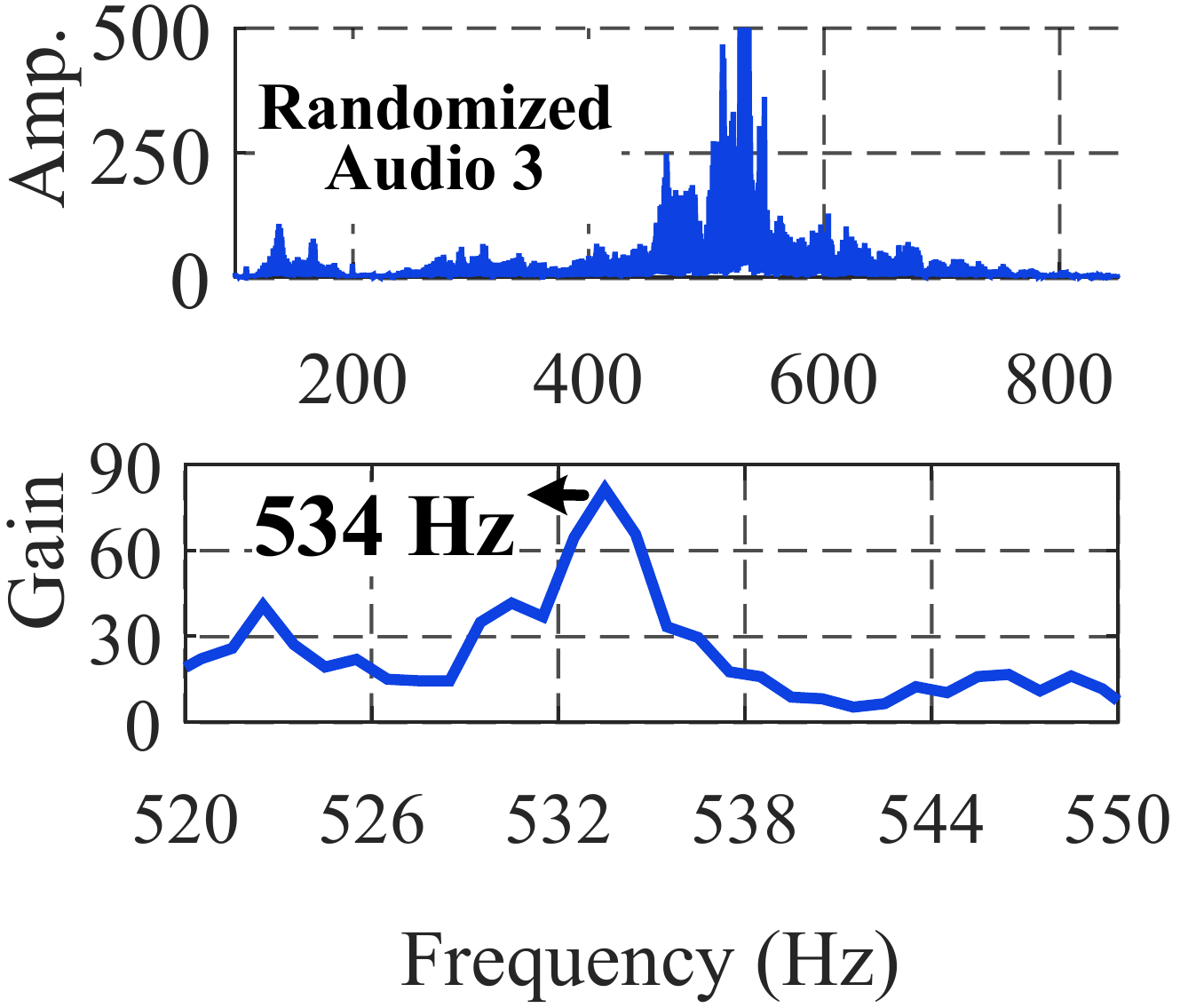}
        \label{R3}}
\caption{Anonymized audio from three randomized structures: (a) audio 1, (b) audio 2, and (c) audio 3.}
\label{RRR}
    \end{minipage}  
\end{figure*}

Speakers naturally change orientation during speech, causing variations in the angle of sound incidence at the microphone. Such variations can degrade anonymization performance in direction-sensitive designs. To address this challenge, we develop a \emph{Dynamically Stable Metamaterial} structure (Sec.~\ref{sec:layout}) and evaluate its effectiveness under realistic dynamic conditions across five ASV models. Fig.~\ref{ST1} and Fig.~\ref{ST2} compare anonymization performance without and with the dynamically stable structure, respectively. With the proposed design, the MMR remains consistently above 90\% across a wide range of incidence angles. In contrast, without the dynamically stable metamaterial, the MMR decreases steadily as the angle deviates and drops to approximately 30\% at $90^\circ$. These results demonstrate that the multi-unit layout is essential for maintaining strong and stable anonymization performance under realistic speaker movement.

\subsubsection{Contribution of complex interference design}\label{B3}
We now evaluate the randomized perturbation mechanism proposed in Sec.~\ref{sec:ril}, which introduces randomness into speech through subtle user movements. In the experiment, the same device was used to repeatedly play the same speech sample three times, while the receiving microphone was slightly moved during each playback, with all other experimental conditions kept identical. We then recorded the corresponding audio spectrograms and computed their gains for comparison~\cite{micpro}. As shown in Fig.~\ref{RRR}, the spectrograms obtained from different playbacks exhibit clear differences, and their gain center frequencies show small variations, confirming the effectiveness of the proposed randomized interference mechanism.


\subsection{Outdoor Experiments\label{sec:prw}}

One of the key advantages of \SystemName is its ability to provide anonymization protection during outdoor speeches or mobile phone calls. To this end, we introduced an outdoor experimental scenario (Fig.~\ref{S22}). In subsequent experiments, we further introduced various environmental interferences to systematically evaluate the method’s anonymization performance and robustness under real-world conditions.

\subsubsection{Impact of mobile environments}\label{C1}
The passive and portable \SystemName can be flexibly integrated with mobile devices such as phones, enabling anonymous remote meetings in dynamic scenarios. In practice, a speaker’s walking speed may affect the propagation path and incidence angle of sound waves, potentially impacting anonymization performance. To evaluate the system’s performance under dynamic conditions, we conducted experiments at different walking speeds and measured the anonymization of speech recorded during movement.

The results show that even at a relatively high walking speed of 2.5 m/s, \SystemName maintains an MMR above 90\% across various ASV models (Fig.~\ref{noise}), demonstrating strong robustness to changes in walking speed. This stability is largely attributed to the Dynamically Stable Metamaterial proposed in Sec.~\ref{sec:layout}, which continuously provides effective physical interference even when the sound incidence angle varies with the speaker’s movement, thereby ensuring that anonymization performance remains unaffected by mobility.

\subsubsection{Impact of environmental noise}\label{C2}
Environmental noise may overlap with speech signals in both spectral and energy distributions, thereby degrading anonymization performance. To evaluate the robustness of \SystemName under noisy conditions, we introduce background white noise at different intensity levels in our experiments and assess its anonymization effectiveness in the presence of noise interference.

\begin{figure}[!t]
\centering
\subfloat[]{
    \includegraphics[scale=0.135]{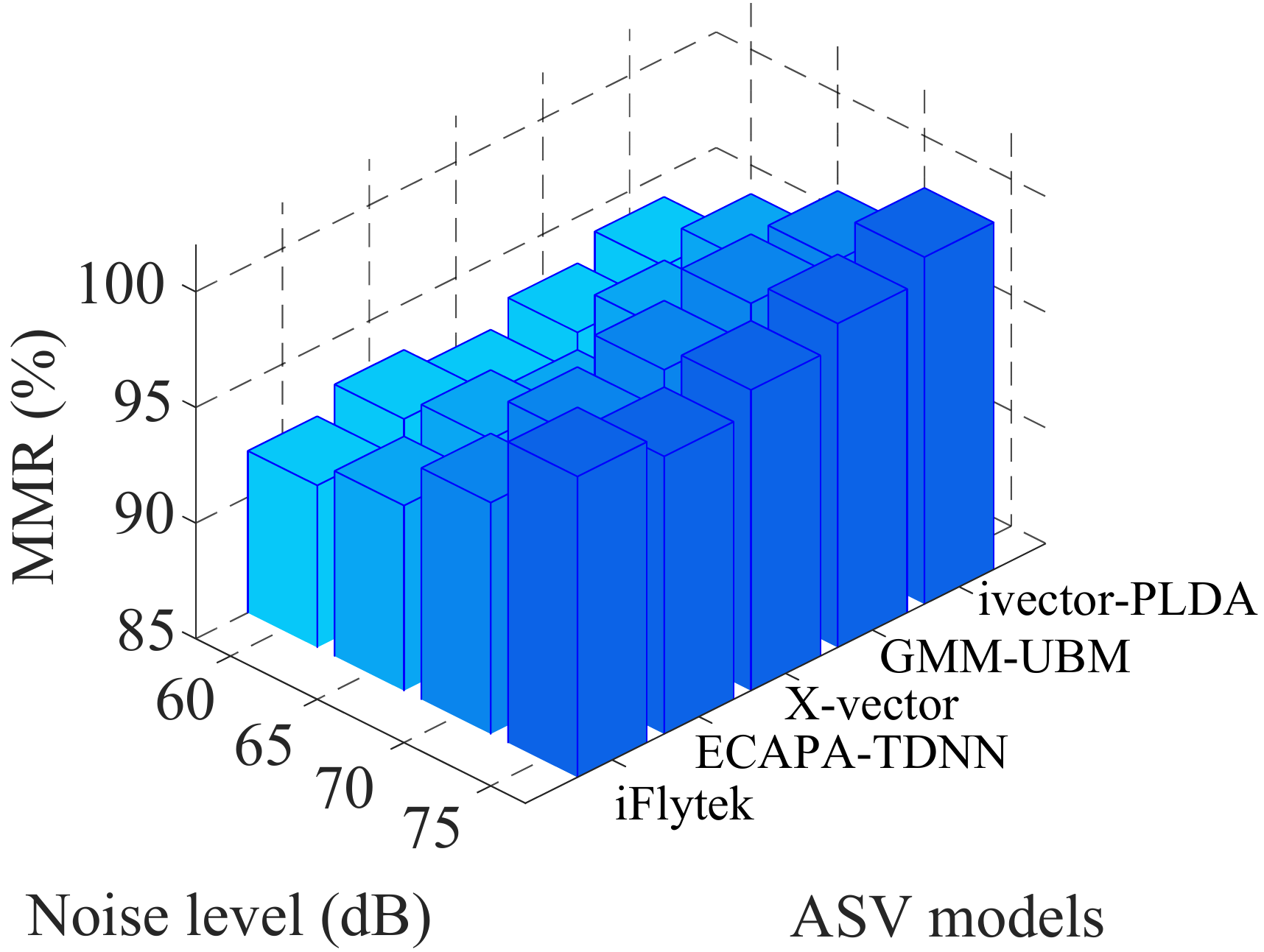}
    \label{noise}}
\hfill
\subfloat[]{
    \includegraphics[scale=0.135]{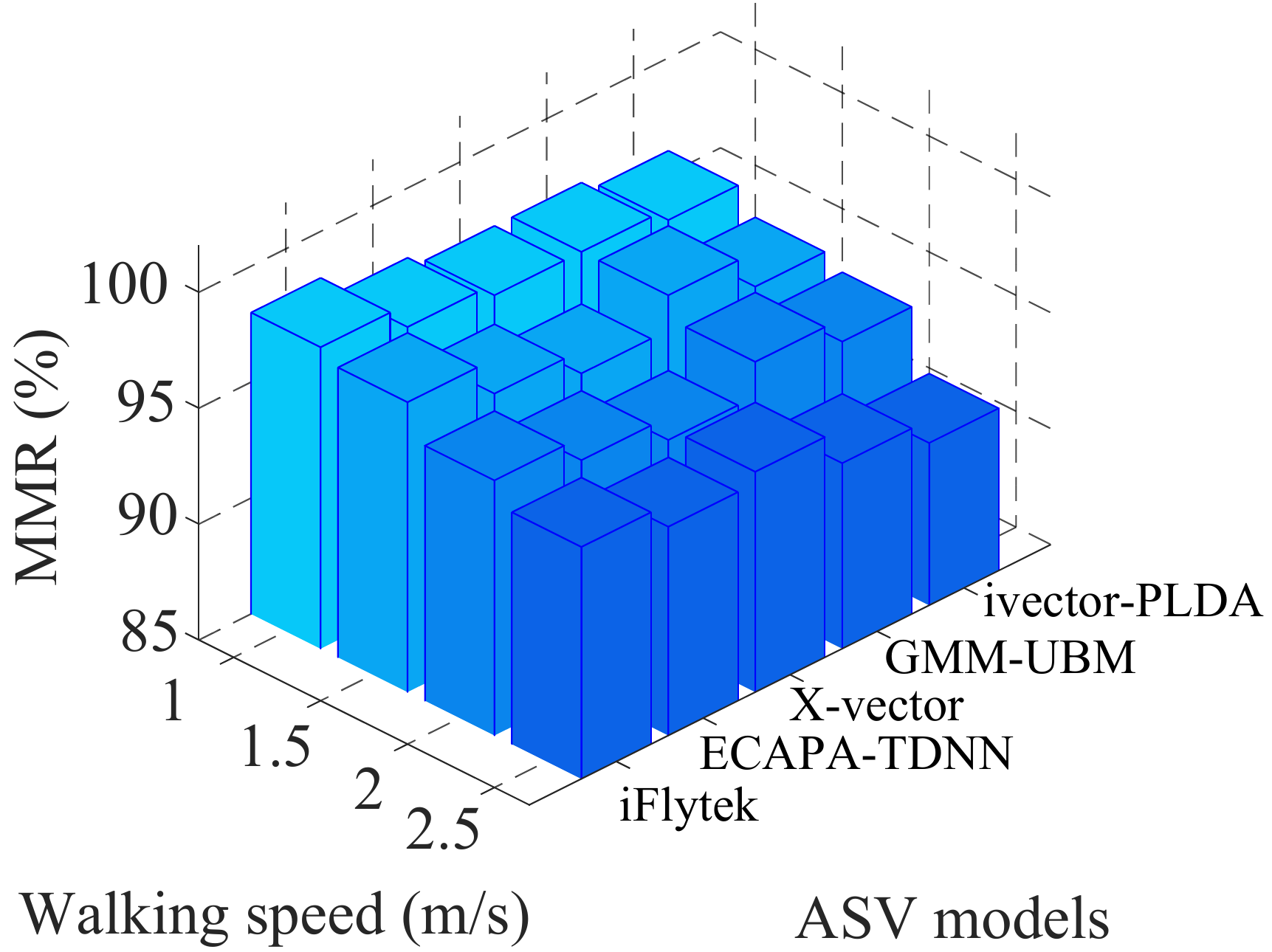}
    \label{move}}
\caption{(a) Anonymization under varying noise levels, (b) under varying walking speeds.}
\label{NaM}
\end{figure}

We evaluated \SystemName under background noise levels ranging from quiet to noisy environments (60–75 dB). As shown in Fig.~\ref{move}, the MMR consistently remains above 90\% across all noise conditions. Notably, anonymization performance improves as noise increases, reaching an average MMR above 97\% at 75 dB. We attribute this effect to the fact that background noise further perturbs speaker-specific cues, while \SystemName's physical-layer interference remains stable and unaffected by noise. The combination of environmental noise and robust physical interference therefore amplifies voiceprint disruption rather than degrading it.

\begin{figure}[t!]
    \centering
    \includegraphics[width=1\linewidth]{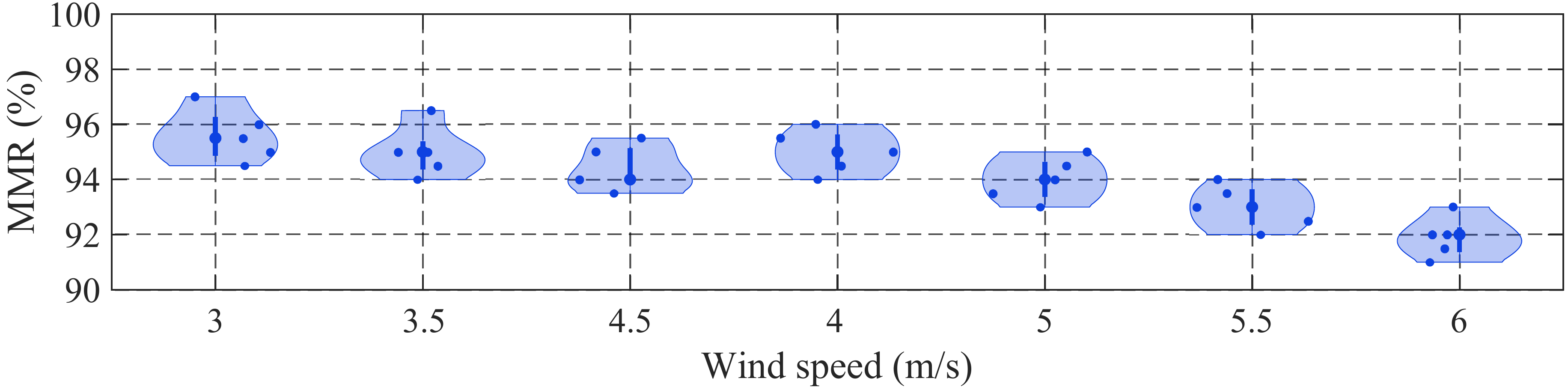}
    \caption{Impact of different wind speeds on \SystemName.}
    \label{windspeed}
\end{figure}
\subsubsection{Impact of wind speed}\label{C3}
When a speaker gives an outdoor speech or participates in a remote meeting, wind can alter the propagation path and attenuation characteristics of sound waves, thereby affecting the interference effect of \SystemName. To this end, we simulated different wind speed conditions in our experiments to evaluate the anonymization performance of \SystemName under wind interference.
Fig.~\ref{windspeed} shows that even at wind speeds of 6 m/s (above the average urban level~\cite{windspeed1,windspeed2}), \SystemName maintains an MMR above 90\%, demonstrating its robustness against wind interference. We attribute this to the multi-unit and multi-angle design of the Dynamically Stable Metamaterial structure, which ensures stable anonymization performance even when sound waves are deflected.

\section{Discussions} \label{chap:9}
\SystemName opens up a new design space for physical-layer voiceprint protection. Naturally, there is room for improvement and further work. We discuss a few points here.

\cparagraph{Enhancing usability and comfort}
While \SystemName already achieves a high level of integration, its current rigid metamaterial units may affect the appearance or form factor of some microphones. An interesting direction for future work is the use of flexible acoustic metamaterials~\cite{spider,spider2}. Their elastic and bendable properties could enable close conformance to diverse microphone shapes, preserving device aesthetics and existing user interaction patterns. Such designs would further improve comfort, deployability, and user acceptance in everyday settings, while remaining compatible with our overall design methodology.

\cparagraph{Enhancing robustness against adaptive attacks}
Our current design employs a narrowband resonant structure with a fixed interference band, which provides strong and efficient anonymization. Looking ahead, this structure can be further strengthened against highly adaptive or targeted attacks through dynamic tuning. For example, incorporating piezoelectric-based metamaterial elements~\cite{piezoelectric1,piezoelectric2} would allow the interference band to be adjusted in real time. Although this extension would require external power supply (e.g., through two AA batteries), it could yield richer, less predictable interference patterns, thereby increasing robustness against spectrum-aware adversaries.

\cparagraph{Optimizing auditory naturalness}
Our evaluation shows that \SystemName preserves speech intelligibility well, but its resonance-based interference can introduce subtle timbre changes. Future work can further refine the perceptual quality of anonymized speech by integrating models of human auditory perception, such as psychoacoustic masking effects~\cite{psychoacoustic1,psychoacoustic2}. By steering interference toward frequency regions that are less perceptible to human listeners yet critical for speaker identification, the system can achieve an even better balance between strong anonymization and natural-sounding speech.

\cparagraph{Evaluation methodology}
While software-based methods rely on large-scale dataset processing, \SystemName, as a deterministic physical system, is evaluated through real acoustic experiments. This distinction arises because our metamaterial's transfer function is fixed by its geometry and material properties, and thus does not require statistical validation on massive datasets to establish its effect. Physical determinism ensures that controlled lab measurements reflect its consistent real-world performance.

\section{Related Work}
\vspace{-2mm}
\begin{table}[t!]
    \scriptsize
    \caption{Comparison with eight prior work}
    \label{COMPARE}
    \vspace{1mm}
    \centering
    \setlength{\tabcolsep}{5pt} 
    \begin{tabular} {p{1.7cm}lllllll}
    \toprule
       \textbf{\makecell[l]{System\\ Name}} & \textbf{\makecell[l]{Source\\ Prot.}} & \textbf{\makecell[l]{No Sys. Res.\\ Required}}  & \textbf{\makecell[l]{No HW\\or SF}} & \textbf{\makecell[l]{Cross Devs.\\ or Models}} & 
    \textbf{\makecell[l]{No\\delay}}
    \\
    \midrule
\rowcolor{gray!20} V-Cloak \cite{vcloak}& No & No & No & Yes   & No\\
VSMask \cite{vsmask} & No & No & No & Yes   & Yes\\
\rowcolor{gray!20} MicPro \cite{micpro}& Yes & Yes & No & No   & Yes\\
EASY \cite{easy} & No & No & No & Yes   & No \\
\rowcolor{gray!20} \makecell[l]{Speech\\ Sanitizer\cite{Speechsanitizer}}  & No & No & No & Yes  & No \\
MUSA \cite{musa} & No & No & No & Yes   & No\\
\rowcolor{gray!20}  Arasteh et al. \cite{natureaddressing}&  No & No & No & Yes   & No\\
Enkidu \cite{enkidu}&  No & No & No & Yes   & Yes\\
\rowcolor{gray!20} \makecell[l]{\textbf{EchoMask}}& \textbf{Yes}& \textbf{Yes}& \textbf{Yes}&\textbf{Yes}  & \textbf{Yes}\\  
    \bottomrule
    \end{tabular}
\end{table}
Our work lies at the intersection of voiceprint anonymization, microphone-level protection, and physical-layer security. Prior efforts have explored both software- and hardware-based defenses to mitigate voiceprint leakage. While effective in controlled settings, these approaches make different assumptions about device trust, system integration, and deployment environments, which limit their applicability in public or shared recording scenarios.

\cparagraph{Software-based solutions}
Software-based approaches anonymize speech after audio captured through signal processing or learning-based transformations to obfuscate speaker identity~\cite{vsmask,vcloak,musa,easy,enkidu,natureaddressing,Speechsanitizer}. 
VSMask~\cite{vsmask} is a learnable masking mechanism that perturbs speaker-discriminative features while preserving speech intelligibility. V-Cloak~\cite{vcloak} and Speech Sanitizer~\cite{Speechsanitizer} explore feature-space transformations to remove identity cues while retaining linguistic content. Enkidu~\cite{enkidu} presents an end-to-end anonymization framework that balances identity removal with the preservation of linguistic and emotional information, enabling fine-grained control over the privacy-utility trade-off.
These systems demonstrate that effective anonymization can be achieved in real time using signal- and model-driven techniques. However, they fundamentally assume that the microphone, firmware, and recording software are trusted. In practice, attackers may intercept or record raw audio before these mechanisms are applied, rendering post-capture anonymization ineffective~\cite{micpro}. Moreover, continuous software processing incurs runtime overhead and may introduce latency, which limits practicality for lightweight or resource-constrained settings.

\cparagraph{Hardware- and microphone-centric defenses}
MicPro~\cite{micpro} represents a hardware-level approach that moves anonymization closer to the audio source by modifying the CELP codec within the microphone pipeline. By reshaping resonance peak features during encoding, MicPro prevents voiceprint leakage before audio leaves the device. Through multi-objective optimization, it balances anonymization strength, speech recognition accuracy, and intelligibility, achieving low-latency protection.
Despite these advantages, MicPro is tightly coupled to CELP-based encoding and microphone hardware limited to an 8\,kHz sampling rate, making it difficult to extend to higher-rate CELP variants or non-CELP audio pipelines~\cite{CELP,CELP2,ACELP,ACELP2}. More broadly, hardware-level solutions often require specialized system integration and device-specific modifications, which significantly constrain deployment in public, shared, or heterogeneous environments.

\cparagraph{Positioning of \SystemName}
Table~\ref{COMPARE} summarizes a comparison between \SystemName and prior work. Unlike software-based approaches, \SystemName does not rely on trusted microphones, firmware, or recording software, and remains effective even if downstream components are compromised. Unlike prior hardware-based solutions, it requires no changes to microphone electronics, codecs, or system pipelines, and is not tied to specific devices or audio formats. By operating solely at the physical layer and modulating sound before capture, \SystemName introduces no system overhead, requires no pre-trained models, and incurs virtually no processing delay. It can be directly deployed on off-the-shelf microphones from different manufacturers, making it particularly well-suited for public and shared recording scenarios where device trust and system control cannot be assumed.





\section{Conclusion} \label{chap:12}
\vspace{-2mm}
We have presented \SystemName, the first physical-layer, power-free approach for protecting voiceprints at the moment of audio capture using acoustic metamaterials. Unlike software-based defenses that operate after digitization, \SystemName prevents identity-bearing information from ever entering the digital pipeline.
\SystemName introduces three core design innovations: targeted interference units derived from spectral differences between voiceprints and speech; a dynamically stable metamaterial layout guided by an anonymized acoustic field model to ensure robust angular coverage; and passive randomization of the acoustic response to improve long-term robustness. Together, these designs enable effective voiceprint protection while preserving speech intelligibility and usability.
Our evaluation shows that \SystemName consistently degrades the accuracy of multiple state-of-the-art speaker recognition systems under identity abuse scenarios. These results demonstrate the viability of passive, physical-layer defenses and point to a new direction for building lightweight and reliable privacy barriers directly at the point of audio capture.

\appendix

\section{Ethical Considerations}
This work investigates a passive, physical-layer voice anonymization technique based on acoustic metamaterials, with the goal of protecting speaker identity while preserving speech usability in everyday settings. We consider the interests of multiple stakeholders, including study participants, end users, device manufacturers, and the broader research community. All user studies were conducted with informed consent, allowed participants to withdraw at any time, and followed data minimization and anonymization principles. No sensitive personal data beyond speech recordings required for evaluation were collected, and all recordings were used solely for research purposes.

We stress that \SystemName is designed to protect against voiceprint capture from compromised or untrusted recording devices and does not aim to prevent all forms of audio recording or law enforcement. While voice anonymization technologies may raise dual-use concerns, the primary goal of this work is to reduce unauthorized speaker identification in public and shared environments. To mitigate misuse, we focus on communicating system principles, feasibility, and performance.
After weighing potential risks against societal benefits, we believe this work contributes positively to privacy-preserving technology development and supports its safe and responsible deployment.




\bibliographystyle{plain}
\bibliography{references}

@String{Computing = "Computing" }

@String{Computer = "{IEEE} Computer" }

@String{Springer = "Springer-Verlag" }

@misc{r1,
title = {Apple-Siri},
author={},
year = {\url{https://openai.com/blog/chatgpt/}},
note={Last accessed: 2024-1-2}
}

@misc{r2,
title = {Amazon-Alexa},
year = {\url{https://developer.amazon.com/alexa}},
note={Last accessed: 2024-1-2}
}

@misc{r3,
title = {HuaWei-Xiaoyi},
year = {\url{https://consumer-img.huawei.com/cn/support/speakers/}},
note={Last accessed: 2024-1-2}
}

@article{r11,
  title={Acoustic metasurfaces},
  author={Assouar, Badreddine and Liang, Bin and Wu, Ying and Li, Yong and Cheng, Jian-Chun and Jing, Yun},
  journal={Nature Reviews Materials},
  volume={3},
  number={12},
  pages={460--472},
  year={2018},
  publisher={Nature Publishing Group UK London}
}

@article{r12,
  title={Acoustic metamaterials: From local resonances to broad horizons},
  author={Ma, Guancong and Sheng, Ping},
  journal={Science advances},
  volume={2},
  number={2},
  pages={e1501595},
  year={2016},
  publisher={American Association for the Advancement of Science}
}

@article{r14,
  title={Nonlocal Ventilating Metasurfaces},
  author={Zhu, Yihuan and Dong, Ruizhi and Mao, Dongxing and Wang, Xu and Li, Yong},
  journal={Physical Review Applied},
  volume={19},
  number={1},
  pages={014067},
  year={2023},
  publisher={APS}
}

@article{r15,
  title={A compact multifunctional metastructure for Low-frequency broadband sound absorption and crash energy dissipation},
  author={Ren, Zhiwen and Cheng, Yuehang and Chen, Mingji and Yuan, Xujin and Fang, Daining},
  journal={Materials \& Design},
  volume={215},
  pages={110462},
  year={2022},
  publisher={Elsevier}
}

@misc{Google,
title = {Google Speech-to-Text AI},
author={},
year = {\url{https://cloud.google.com/speech-to-text}},
note={Last accessed: 2024-8-6}
}

@misc{Googlep,
title = {Google},
author={},
year = {\url{https://www.google-mobile.cn/}},
note={Last accessed: 2025-1-20}
}

@misc{Huawei,
title = {Huawei},
author={},
year = {\url{https://consumer.huawei.com/cn/phones/}},
note={Last accessed: 2025-1-20}
}

@inproceedings{windspeed1,
  title={Analysis of temperature, air humidity and wind conditions for the needs of outdoor thermal comfort},
  author={Dec, Ewelina and Babiarz, Bo{\.z}ena and Sekret, Robert},
  booktitle={E3S Web of Conferences},
  volume={44},
  pages={00028},
  year={2018},
  organization={EDP Sciences}
}

@article{windspeed2,
  title={Numerical evaluation of urban geometry's control of wind movements in outdoor spaces during winter period. Case of Mediterranean climate},
  author={Bouketta, S and Bouchahm, Y},
  journal={Renewable Energy},
  volume={146},
  pages={1062--1069},
  year={2020},
  publisher={Elsevier}
}

@article{mie,
  title={Ultra-sparse metasurface for high reflection of low-frequency sound based on artificial Mie resonances},
  author={Cheng, Y and Zhou, C and Yuan, BG and Wu, DJ and Wei, Q and Liu, XJ},
  journal={Nature materials},
  volume={14},
  number={10},
  pages={1013--1019},
  year={2015},
  publisher={Nature Publishing Group UK London}
}

@article{mitigating,
  title={Mitigating inaudible ultrasound attacks on voice assistants with acoustic metamaterials},
  author={Lloyd, Joshua S and Ludwikowski, Cole G and Malik, Cyrus and Shen, Chen},
  journal={IEEE Access},
  volume={11},
  pages={36464--36470},
  year={2023},
  publisher={IEEE}
}

@inproceedings{vsmask,
  title={Vsmask: Defending against voice synthesis attack via real-time predictive perturbation},
  author={Wang, Yuanda and Guo, Hanqing and Wang, Guangjing and Chen, Bocheng and Yan, Qiben},
  booktitle={Proceedings of the 16th ACM Conference on Security and Privacy in Wireless and Mobile Networks},
  pages={239--250},
  year={2023}
}

@article{musa,
  title={MUSA: Multi-lingual speaker anonymization via serial disentanglement},
  author={Yao, Jixun and Wang, Qing and Guo, Pengcheng and Ning, Ziqian and Yang, Yuguang and Pan, Yu and Xie, Lei},
  journal={IEEE Transactions on Audio, Speech and Language Processing},
  year={2025},
  publisher={IEEE}
}

@inproceedings{vcloak,
  title={$\{$V-Cloak$\}$: Intelligibility-, naturalness-\& $\{$Timbre-Preserving$\}$$\{$Real-Time$\}$ voice anonymization},
  author={Deng, Jiangyi and Teng, Fei and Chen, Yanjiao and Chen, Xiaofu and Wang, Zhaohui and Xu, Wenyuan},
  booktitle={32nd USENIX Security Symposium (USENIX Security 23)},
  pages={5181--5198},
  year={2023}
}

@inproceedings{micpro,
  title={MicPro: Microphone-based voice privacy protection},
  author={Xiao, Shilin and Ji, Xiaoyu and Yan, Chen and Zheng, Zhicong and Xu, Wenyuan},
  booktitle={Proceedings of the 2023 ACM SIGSAC Conference on Computer and Communications Security},
  pages={1302--1316},
  year={2023}
}

@article{Voiceprint1,
  title={Voiceprint recognition technology and its application status},
  author={Zheng, Fang and Li, LT and Zhang, Hui},
  journal={Research on information security},
  volume={2},
  number={1},
  pages={44--57},
  year={2016}
}

@inproceedings{Voiceprint2,
  title={LV-auth: Lip Motion Fusion for Voiceprint Authentication},
  author={Liu, Wei and Zhu, Xiaojing and Liu, Qin and Li, Peng and Zhou, Man},
  booktitle={International Conference on Wireless Artificial Intelligent Computing Systems and Applications},
  pages={295--307},
  year={2024},
  organization={Springer}
}

@inproceedings{Voiceprint3,
  title={Security system using biometric technology: Design and implementation of Voice Recognition System (VRS)},
  author={Rashid, Rozeha A and Mahalin, Nur Hija and Sarijari, Mohd Adib and Aziz, Ahmad Aizuddin Abdul},
  booktitle={2008 international conference on computer and communication engineering},
  pages={898--902},
  year={2008},
  organization={IEEE}
}

@inproceedings{Voiceprint4,
  title={Voiceprint-biometric template design and authentication based on cloud computing security},
  author={Zhu, Hua-Hong and He, Qian-Hua and Tang, Hong and Cao, Wei-Hua},
  booktitle={2011 International Conference on Cloud and Service Computing},
  pages={302--308},
  year={2011},
  organization={IEEE}
}

@inproceedings{Voiceprint5,
  title={Voice biometrics: Deep learning-based voiceprint authentication system},
  author={Boles, Andrew and Rad, Paul},
  booktitle={2017 12th system of systems engineering conference (SoSE)},
  pages={1--6},
  year={2017},
  organization={IEEE}
}

@inproceedings{attack1,
  title={Your voice assistant is mine: How to abuse speakers to steal information and control your phone},
  author={Diao, Wenrui and Liu, Xiangyu and Zhou, Zhe and Zhang, Kehuan},
  booktitle={Proceedings of the 4th ACM workshop on security and privacy in smartphones \& mobile devices},
  pages={63--74},
  year={2014}
}

@inproceedings{attack2,
  title={Real-time, universal, and robust adversarial attacks against speaker recognition systems},
  author={Xie, Yi and Shi, Cong and Li, Zhuohang and Liu, Jian and Chen, Yingying and Yuan, Bo},
  booktitle={ICASSP 2020-2020 IEEE international conference on acoustics, speech and signal processing (ICASSP)},
  pages={1738--1742},
  year={2020},
  organization={IEEE}
}

@inproceedings{attack3,
  title={Imperio: Robust over-the-air adversarial examples for automatic speech recognition systems},
  author={Sch{\"o}nherr, Lea and Eisenhofer, Thorsten and Zeiler, Steffen and Holz, Thorsten and Kolossa, Dorothea},
  booktitle={Proceedings of the 36th Annual Computer Security Applications Conference},
  pages={843--855},
  year={2020}
}

@inproceedings{attack4,
  title={Alexa versus alexa: Controlling smart speakers by self-issuing voice commands},
  author={Esposito, Sergio and Sgandurra, Daniele and Bella, Giampaolo},
  booktitle={Proceedings of the 2022 ACM on Asia Conference on Computer and Communications Security},
  pages={1064--1078},
  year={2022}
}

@inproceedings{attack5,
  title={Practical adversarial attacks against speaker recognition systems},
  author={Li, Zhuohang and Shi, Cong and Xie, Yi and Liu, Jian and Yuan, Bo and Chen, Yingying},
  booktitle={Proceedings of the 21st international workshop on mobile computing systems and applications},
  pages={9--14},
  year={2020}
}

@inproceedings{enkidu,
  title={Enkidu: Universal frequential perturbation for real-time audio privacy protection against voice deepfakes},
  author={Feng, Zhou and Chen, Jiahao and Zhou, Chunyi and Pu, Yuwen and Li, Qingming and Du, Tianyu and Ji, Shouling},
  booktitle={Proceedings of the 33rd ACM International Conference on Multimedia},
  pages={11638--11647},
  year={2025}
}

@article{NTU-NPU,
  title={NTU-NPU System for Voice Privacy 2024 Challenge},
  author={Kuzmin, Nikita and Luong, Hieu-Thi and Yao, Jixun and Xie, Lei and Lee, Kong Aik and Chng, Eng Siong},
  journal={emotion},
  volume={1},
  pages={2}
}

@incollection{saic,
  title={Saic: Integration of speech anonymization and identity classification},
  author={Cheng, Ming and Diao, Xingjian and Cheng, Shitong and Liu, Wenjun},
  booktitle={AI for Health Equity and Fairness: Leveraging AI to Address Social Determinants of Health},
  pages={295--306},
  year={2024},
  publisher={Springer}
}

@article{mie2,
  title={Deep sub-wavelength acoustic transmission enhancement and whisper via the monopole resonance in meta-cavities},
  author={Lei, Yunzhong and Wu, Jiu Hui and Wang, Libo and Huang, Yao and Niu, Jiamin},
  journal={Applied Acoustics},
  volume={203},
  pages={109227},
  year={2023},
  publisher={Elsevier}
}

@article{nature,
  title={Remote whispering metamaterial for non-radiative transceiving of ultra-weak sound},
  author={Zhang, Jin and Rui, Wei and Ma, Chengrong and Cheng, Ying and Liu, Xiaojun and Christensen, Johan},
  journal={Nature Communications},
  volume={12},
  number={1},
  pages={3670},
  year={2021},
  publisher={Nature Publishing Group UK London}
}

@inproceedings{easy,
  title     = {{EASY: Emotion-aware Speaker Anonymization via Factorized Distillation}},
  author    = {Jixun Yao and Hexin Liu and Eng Siong Chng and Lei Xie},
  year      = {2025},
  booktitle = {{Interspeech 2025}},
  pages     = {3219--3223},
  doi       = {10.21437/Interspeech.2025-194},
  issn      = {2958-1796},
}

@inproceedings{songbsab,
  title={Songbsab: A dual prevention approach against singing voice conversion based illegal song covers},
  author={Chen, Guangke and Zhang, Yedi},
  booktitle={32nd Annual Network and Distributed System Security Symposium},
  year={2025}
}

@article{navigating1,
  title={Navigating the tradeoff between personal privacy and data utility in speech anonymization for clinical research},
  author={Diaz-Asper, Catherine and Bongo, Lars Ailo and Elvev{\aa}g, Brita},
  journal={npj Digital Medicine},
  volume={8},
  number={1},
  pages={616},
  year={2025},
  publisher={Nature Publishing Group UK London}
}

@article{navigating2,
  title={Differential privacy enables fair and accurate AI-based analysis of speech disorders while protecting patient data},
  author={Tayebi Arasteh, Soroosh and Lotfinia, Mahshad and Perez-Toro, Paula Andrea and Arias-Vergara, Tomas and Ranji, Mahtab and Orozco-Arroyave, Juan Rafael and Schuster, Maria and Maier, Andreas and Yang, Seung Hee},
  journal={npj Artificial Intelligence},
  volume={1},
  number={1},
  pages={37},
  year={2025},
  publisher={Nature Publishing Group UK London}
}

@article{CELP,
  title={A 16 kb/s wideband CELP-based speech coder using mel-generalized cepstral analysis},
  author={Koishida, Kazuhito and Hirabayashi, Gou and Tokuda, Keiichi and Kobayashi, Takao},
  journal={IEICE transactions on information and systems},
  volume={83},
  number={4},
  pages={876--883},
  year={2000},
  publisher={The Institute of Electronics, Information and Communication Engineers}
}

@inproceedings{CELP2,
  title={A wideband CELP speech coder at 16 kbit/s based on mel-generalized cepstral analysis},
  author={Koishida, Kazuhito and Hirabayashi, Gou and Tokuda, Keiichi and Kobayashi, Takao},
  booktitle={Proceedings of the 1998 IEEE International Conference on Acoustics, Speech and Signal Processing, ICASSP'98 (Cat. No. 98CH36181)},
  volume={1},
  pages={161--164},
  year={1998},
  organization={IEEE}
}

@inproceedings{ACELP,
  title={Techniques for high-quality ACELP coding of wideband speech.},
  author={Bessette, Bruno and Lefebvre, Roch and Salami, Redwan and Jelinek, Milan and Vainio, Janne and Rotola-Pukkila, J and Mikkola, Hannu and J{\"a}rvinen, Kari},
  booktitle={INTERSPEECH},
  pages={1997--2000},
  year={2001}
}

@inproceedings{ACELP2,
  title={A 13.0 kbit/s wideband speech codec based on SB-ACELP},
  author={Schnitzler, J{\"u}rgen},
  booktitle={Proceedings of the 1998 IEEE International Conference on Acoustics, Speech and Signal Processing, ICASSP'98 (Cat. No. 98CH36181)},
  volume={1},
  pages={157--160},
  year={1998},
  organization={IEEE}
}

@inproceedings{micattack1,
author = {Deng, Jiangyi and Chen, Yanjiao and Xu, Wenyuan},
title = {FenceSitter: Black-box, Content-Agnostic, and Synchronization-Free Enrollment-Phase Attacks on Speaker Recognition Systems},
year = {2022},
isbn = {9781450394505},
publisher = {Association for Computing Machinery},
address = {New York, NY, USA},
url = {https://doi.org/10.1145/3548606.3559357},
doi = {10.1145/3548606.3559357},
booktitle = {Proceedings of the 2022 ACM SIGSAC Conference on Computer and Communications Security},
pages = {755–767},
numpages = {13},
location = {Los Angeles, CA, USA},
series = {CCS '22}
}

@article{micattack2,
  title={Electromagnetic interference attacks on cyber-physical systems: Theory, demonstration, and defense},
  author={Dayanikli, Gokcen Yilmaz},
  year={2021},
  publisher={Virginia Tech}
}

@article{micattack3,
  title={Vulnerability of MEMS gyroscopes to targeted acoustic attacks},
  author={Khazaaleh, Shadi and Korres, Georgios and Eid, Mohammed and Rasras, Mahmoud and Daqaq, Mohammed F},
  journal={IEEE Access},
  volume={7},
  pages={89534--89543},
  year={2019},
  publisher={IEEE}
}

@article{micattack4,
  title={Silicon listening. MEMS, near-ultrasound, and machine listening beyond AI},
  author={D{\"o}rfling, Christina},
  journal={Sound Studies},
  volume={11},
  number={2},
  pages={314--338},
  year={2025},
  publisher={Taylor \& Francis}
}

@article{spider,
  title={Spider web-inspired acoustic metamaterials with multi-band gaps for low-frequency elastic wave propagation control},
  author={Wang, Yang and Wang, Xiaoyu and Dong, Huanyu and Ma, Lele and Fu, Yue and Zhang, Lingxing},
  journal={Journal of Physics D: Applied Physics},
  year={2025}
}

@article{spider2,
  title={Development and Characterization of a Flexible Soundproofing Metapanel for Noise Reduction},
  author={Dongil, Jang and Sanha, Kang and Jinyoung, Kim and Hyeonghoon, Kim and Sinwoo, Lee and Bongjoong, Kim},
  year={2024},
  publisher={MDPI}
}

@article{piezoelectric1,
  title={Artificial piezoelectric metamaterials},
  author={Gao, Ziyan and Lei, Yu and Li, Zhanmiao and Yang, Jikun and Yu, Bo and Yuan, Xiaoting and Hou, Zewei and Hong, Jiawang and Dong, Shuxiang},
  journal={Progress in Materials Science},
  pages={101434},
  year={2025},
  publisher={Elsevier}
}

@article{piezoelectric2,
  title={Architected cellular piezoelectric metamaterials: Thermo-electro-mechanical properties},
  author={Shi, Jiahao and Akbarzadeh, AH},
  journal={Acta Materialia},
  volume={163},
  pages={91--121},
  year={2019},
  publisher={Elsevier}
}

@article{psychoacoustic1,
  title={Integrated psychoacoustic active noise control and masking},
  author={Belyi, Valiantsin and Gan, Woon-Seng},
  journal={Applied Acoustics},
  volume={145},
  pages={339--348},
  year={2019},
  publisher={Elsevier}
}

@inproceedings{psychoacoustic2,
  title={Real-time psychoacoustic frequency masking compensation for audio signals with overlapping spectra},
  author={Presti, Giorgio and Degiorgi, Nicola and Fresia, Amedeo and Servetti, Antonio and others},
  booktitle={Proceedings of the 21st Sound and Music Computing Conference},
  pages={439--444},
  year={2024},
  organization={SMC}
}

@article{head1,
  title={The contribution of head movement to the externalization and internalization of sounds},
  author={Brimijoin, W Owen and Boyd, Alan W and Akeroyd, Michael A},
  journal={PloS one},
  volume={8},
  number={12},
  pages={e83068},
  year={2013},
  publisher={Public Library of Science San Francisco, USA}
}

@techreport{head2,
  title={3-D sound for virtual reality and multimedia},
  author={Begault, Durand R and Trejo, Leonard J},
  year={2000}
}

@misc{comsol,
title = {COMSOL},
author={},
year = {\url{https://www.comsol.com/}},
note={Last accessed: 2025-1-20}
}

@article{interference1,
  title={Ultra-open acoustic metamaterial silencer based on Fano-like interference},
  author={Ghaffarivardavagh, Reza and Nikolajczyk, Jacob and Anderson, Stephan and Zhang, Xin},
  journal={Physical Review B},
  volume={99},
  number={2},
  pages={024302},
  year={2019},
  publisher={APS}
}

@article{interference2,
  title={Shaping reverberating sound fields with an actively tunable metasurface},
  author={Ma, Guancong and Fan, Xiying and Sheng, Ping and Fink, Mathias},
  journal={Proceedings of the National Academy of Sciences},
  volume={115},
  number={26},
  pages={6638--6643},
  year={2018},
  publisher={National Academy of Sciences}
}

@article{sounds1,
  title={Mouth sounds: A review of Acoustic Applications and methodologies},
  author={Naal-Ruiz, Norberto E and Gonzalez-Rodriguez, Erick A and Navas-Reascos, Gustavo and Romo-De Leon, Rebeca and Solorio, Alejandro and Alonso-Valerdi, Luz M and Ibarra-Zarate, David I},
  journal={Applied Sciences},
  volume={13},
  number={7},
  pages={4331},
  year={2023},
  publisher={MDPI}
}

@inproceedings{gooseee,
  title={Mobile 3D augmented-reality system for ultrasound applications},
  author={Palmer, Cameron Lowell and Haugen, Bj{\o}rn Olav and Tegnander, Eva and Eik-Nes, Sturla H and Torp, Hans and Kiss, Gabriel},
  booktitle={2015 IEEE International Ultrasonics Symposium (IUS)},
  pages={1--4},
  year={2015},
  organization={IEEE}
}

@article{headmove1,
  title={Head movements while recognizing speech arriving from behind},
  author={Shen, Yi and Folkerts, Monica L and Richards, Virgina M},
  journal={The Journal of the Acoustical Society of America},
  volume={141},
  number={2},
  pages={EL108--EL114},
  year={2017},
  publisher={AIP Publishing}
}

@article{headmove2,
  title={Head-orienting behaviors during simultaneous speech detection and localization},
  author={Lertpoompunya, Angkana and Ozmeral, Erol J and Higgins, Nathan C and Eddins, David A},
  journal={Frontiers in Psychology},
  volume={15},
  pages={1425972},
  year={2024},
  publisher={Frontiers Media SA}
}

@misc{iFlytekASV,
title = {iFlytek},
author={},
year = {\url{https://console.xfyun.cn/services/ivp}},
note={Last accessed: 2026-2-01}
}

@misc{ECAPA-TDNN,
title = {ECAPA-TDNN},
author={},
year = {\url{https://github.com/TaoRuijie/ECAPA-TDNN}},
note={Last accessed: 2026-2-01}
}

@inproceedings{Xvectors,
  title={X-vectors: Robust dnn embeddings for speaker recognition},
  author={Snyder, David and Garcia-Romero, Daniel and Sell, Gregory and Povey, Daniel and Khudanpur, Sanjeev},
  booktitle={2018 IEEE international conference on acoustics, speech and signal processing (ICASSP)},
  pages={5329--5333},
  year={2018},
  organization={IEEE}
}

@article{ivectorPLDA ,
  title={Front-end factor analysis for speaker verification},
  author={Dehak, Najim and Kenny, Patrick J and Dehak, R{\'e}da and Dumouchel, Pierre and Ouellet, Pierre},
  journal={IEEE Transactions on Audio, Speech, and Language Processing},
  volume={19},
  number={4},
  pages={788--798},
  year={2010},
  publisher={IEEE}
}

@article{GMMUBM,
  title={Speaker verification using adapted Gaussian mixture models},
  author={Reynolds, Douglas A and Quatieri, Thomas F and Dunn, Robert B},
  journal={Digital signal processing},
  volume={10},
  number={1-3},
  pages={19--41},
  year={2000},
  publisher={Elsevier}
}

@misc{Shuresv200,
title = {Shure SV200},
author={},
year = {\url{https://www.shure.com/en-ASIA/products/microphones/sv200}},
note={Last accessed: 2026-2-01}
}

@misc{BehringerTA5212,
title = {Behringer TA5212},
author={},
year = {\url{https://www.sweelee.com.sg/products/behringer-ta5212-condenser-gooseneck-microphone}},
note={Last accessed: 2026-2-01}
}

@misc{Audio-TechnicaAT9930,
title = {Audio-Technica AT9930},
author={},
year = {\url{https://www.audio-technica.com.hk/index.php?op=productdetails&pid=478&lang=eng}},
note={Last accessed: 2026-2-01}
}

@misc{sEElectronicsV7 ,
title = {sE Electronics V7},
author={},
year = {\url{https://seelectronics.com/products/v7/}},
note={Last accessed: 2026-2-01}
}

@article{Speechsanitizer,
  title={Speech sanitizer: Speech content desensitization and voice anonymization},
  author={Qian, Jianwei and Du, Haohua and Hou, Jiahui and Chen, Linlin and Jung, Taeho and Li, Xiang-Yang},
  journal={IEEE Transactions on Dependable and Secure Computing},
  volume={18},
  number={6},
  pages={2631--2642},
  year={2019},
  publisher={IEEE}
}

@article{natureaddressing,
  title={Addressing challenges in speaker anonymization to maintain utility while ensuring privacy of pathological speech},
  author={Tayebi Arasteh, Soroosh and Arias-Vergara, Tom{\'a}s and P{\'e}rez-Toro, Paula Andrea and Weise, Tobias and Packh{\"a}user, Kai and Schuster, Maria and Noeth, Elmar and Maier, Andreas and Yang, Seung Hee},
  journal={Communications Medicine},
  volume={4},
  number={1},
  pages={182},
  year={2024},
  publisher={Nature Publishing Group UK London}
}

@inproceedings{SP2025,
  title={From one stolen utterance: Assessing the risks of voice cloning in the aigc era},
  author={Wang, Kun and Chen, Meng and Lu, Li and Feng, Jingwen and Chen, Qianniu and Ba, Zhongjie and Ren, Kui and Chen, Chun},
  booktitle={2025 IEEE Symposium on Security and Privacy (SP)},
  pages={4663--4681},
  year={2025},
  organization={IEEE}
}

@article{speakerrecognition1,
  title={Robust speaker recognition using denoised vocal source and vocal tract features},
  author={Wang, Ning and Ching, PC and Zheng, Nengheng and Lee, Tan},
  journal={IEEE transactions on audio, speech, and language processing},
  volume={19},
  number={1},
  pages={196--205},
  year={2010},
  publisher={IEEE}
}

@inproceedings{speakerrecognition2,
  title={Speaker Identity and Voice Quality: Modeling Human Responses and Automatic Speaker Recognition.},
  author={Park, Soo Jin and Sigouin, Caroline and Kreiman, Jody and Keating, Patricia A and Guo, Jinxi and Yeung, Gary and Kuo, Fang-Yu and Alwan, Abeer},
  booktitle={Interspeech},
  pages={1044--1048},
  year={2016}
}

@article{formant2,
  title={Effects of the rate of formant-frequency variation on the grouping of formants in speech perception},
  author={Summers, Robert J and Bailey, Peter J and Roberts, Brian},
  journal={Journal of the Association for Research in Otolaryngology},
  volume={13},
  number={2},
  pages={269--280},
  year={2012},
  publisher={Springer}
}

@article{formant1,
  title={Text-independent speaker identification using vowel formants},
  author={Almaadeed, Noor and Aggoun, Amar and Amira, Abbes},
  journal={Journal of Signal Processing Systems},
  volume={82},
  number={3},
  pages={345--356},
  year={2016},
  publisher={Springer}
}

@article{MFCC,
  title={Mel frequency cepstral coefficient: a review},
  author={Ali, Shalbbya and Tanweer, Safdar and Khalid, Syed Sibtain and Rao, Naseem},
  journal={ICIDSSD},
  year={2020}
}

@article{attacks1,
  title={Introducing model inversion attacks on automatic speaker recognition},
  author={Pizzi, Karla and Boenisch, Franziska and Sahin, Ugur and B{\"o}ttinger, Konstantin},
  journal={arXiv preprint arXiv:2301.03206},
  year={2023}
}

@article{attacks2,
  title={Privacy leakage on dnns: A survey of model inversion attacks and defenses},
  author={Fang, Hao and Qiu, Yixiang and Yu, Hongyao and Yu, Wenbo and Kong, Jiawei and Chong, Baoli and Chen, Bin and Wang, Xuan and Xia, Shu-Tao and Xu, Ke},
  journal={arXiv preprint arXiv:2402.04013},
  year={2024}
}

@misc{Apple,
title = {Apple},
author={},
year = {\url{https://www.apple.com.cn/iphone/}},
note={Last accessed: 2026-1-14}
}

@misc{samsungS24,
title = {Samsung},
author={},
year = {\url{https://www.samsung.com/hk/smartphones/galaxy-s24/}},
note={Last accessed: 2026-1-14}
}

@article{asr1,
  title={ASR-based speech intelligibility prediction: A review},
  author={Karbasi, Mahdie and Kolossa, Dorothea},
  journal={Hearing Research},
  volume={426},
  pages={108606},
  year={2022},
  publisher={Elsevier}
}

@inproceedings{asr2,
  title={Automatic speech recognition (ASR) based approach for speech therapy of aphasic patients: A review},
  author={Jamal, Norezmi and Shanta, Shahnoor and Mahmud, Farhanahani and Sha’abani, MNAH},
  booktitle={AIP Conference Proceedings},
  volume={1883},
  number={1},
  pages={020028},
  year={2017},
  organization={AIP Publishing LLC}
}

@inproceedings{impact1,
  title={The effect of the frequency and energetic content of broadband noise on the lombard effect and speech intelligibility},
  author={Bottalico, Pasquale and Murgia, Silvia},
  booktitle={Acoustics},
  volume={5},
  number={4},
  pages={898--908},
  year={2023},
  organization={MDPI}
}
\end{document}